\documentclass{ptephy_v1}
\usepackage{graphicx}
\usepackage{subfigure}
\usepackage{multicol}             
\usepackage{multirow}
\usepackage{here}
\usepackage{lineno}

\preprintnumber{XXXX-XXXX} 

\definecolor{green(html/cssgreen)}{rgb}{0.0, 0.5, 0.0}

\begin{document}

\title{First measurement using a nuclear emulsion detector of the $\nu_{\mu}$ charged\nobreakdash-current cross section on iron around the 1\,GeV energy region}

\author[1]{H.~Oshima}
\affil{Toho University, Department of Physics, Funabashi 274--8510, Japan} 
\author[1]{T.~Matsuo\thanks{Present address: Nagoya University.}}
\author{A.~Ali}
\affil{Kyoto University, Department of Physics, Kyoto 606--8502, Japan} 
\author{S.~Aoki}
\affil{Kobe University, Kobe 657--8501, Japan} 
\author{L.~Berns}
\affil{Tokyo Institute of Technology, Department of Physics, Tokyo 152--8551, Japan} 
\author{T.~Fukuda}
\affil{Nagoya University, Nagoya 464--8602, Japan} 
\author{Y.~Hanaoka}
\affil{Nihon University, Narashino 275--8576, Japan} 
\author{Y.~Hayato}
\affil{University of Tokyo, Institute for Cosmic Ray Research, Kamioka Observatory, Kamioka 506--1205, Japan} 
\author[2]{A.~Hiramoto}
\author[2]{A.~K.~Ichikawa}
\author[5]{H.~Kawahara}
\author[2]{T.~Kikawa}
\author[5]{R.~Komatani}
\author[5]{M.~Komatsu}
\author[3]{K.~Kuretsubo}
\author[3]{T.~Marushima}
\author[3]{H.~Matsumoto}
\author[6]{S.~Mikado}
\author{A.~Minamino}
\affil{Yokohama National University, Yokohama 240--8501, Japan} 
\author[1]{K.~Mizuno}
\author[1]{Y.~Morimoto}
\author[5]{K.~Morishima}
\author[5]{N.~Naganawa}
\author[5]{M.~Naiki}
\author[5]{M.~Nakamura}
\author[5]{Y.~Nakamura}
\author[5]{N.~Nakano}
\author[5]{T.~Nakano}
\author[2]{T.~Nakaya}
\author[5]{A.~Nishio}
\author[2]{T.~Odagawa}
\author[1, *]{S.~Ogawa}
\author[5]{H.~Rokujo}
\author[5]{O.~Sato}
\author[1]{H.~Shibuya}
\author[5]{K.~Sugimura}
\author[5]{L.~Suzui}
\author[5]{Y.~Suzuki}
\author[1]{H.~Takagi}
\author[3]{S.~Takahashi}
\author[5]{T.~Takao}
\author[8]{Y.~Tanihara}
\author[5]{R.~Watanabe}
\author[3]{K.~Yamada}
\author[2]{K.~Yasutome}
\author{M.~Yokoyama}
\affil{University of Tokyo, Department of Physics, Tokyo 113--0033, Japan \email{ogawa@ph.sci.toho-u.ac.jp}} 

\begin{abstract}
We have carried out $\nu_{\mu}$ charged\nobreakdash-current interaction measurement on iron using an emulsion detector exposed to the T2K neutrino beam in the J\nobreakdash-PARC neutrino facility.
The data samples correspond to 4.0$\times$10$^{19}$ protons on target, and the neutrino mean energy is 1.49\,GeV.
The emulsion detector is suitable for precision measurements of charged particles produced in neutrino-iron interactions with a low momentum threshold thanks to thin-layered structure and sub-$\mu$m spatial resolution.
The charged particles are successfully detected, and their multiplicities are measured using the emulsion detector.
The cross section was measured to be $\sigma^{\mathrm{Fe}}_{\mathrm{CC}} =  (1.28 \pm 0.11({\mathrm{stat.}})^{+0.12}_{-0.11}({\mathrm{syst.}})) \times 10^{-38} \,  {\mathrm{cm}}^{2}/{\mathrm{nucleon}}$.
The cross section in a limited kinematic phase space of induced muons, $\theta_{\mu} < 45^{\circ}$ and $p_{\mu} > 400 \, {\rm MeV}/c$, on iron was $\sigma^{\mathrm{Fe}}_{\mathrm{CC \hspace{1mm} phase \hspace{0.5mm} space}} = (0.84 \pm 0.07({\mathrm{stat.}})^{+0.07}_{-0.06}({\mathrm{syst.}})) \times 10^{-38} \,  {\mathrm{cm}}^{2}/{\mathrm{nucleon}}$.
The cross-section results are consistent with previous values obtained via different techniques using the same beamline, and they are well reproduced by current neutrino interaction models.
These results demonstrate the capability of the detector towards the detailed measurements of the neutrino-nucleus interactions around the 1\,GeV energy region.
\end{abstract}

\subjectindex{C32}

\maketitle


\section{Introduction}\label{sec:INTRODUCTION}
Long-baseline neutrino oscillation experiments, such as T2K~\cite{T2K_2011}, are typically performed in the vicinity of 1\,GeV.
It is essential to understand neutrino-nucleus interactions for future neutrino oscillation experiments because the experimental precision will be limited by uncertainties of neutrino interaction models.
In this energy region, the dominant modes of neutrino charged\nobreakdash-current interactions are quasi-elastic scattering and resonant pion production.
In addition, the existence of two-particle-two-hole excitations has been posited.
Measuring the multiplicity and kinematics of protons and pions from neutrino interactions is important for constructing reliable neutrino interaction models.
It is a difficult task, however, since the produced hadrons have low energies.
It is especially difficult to observe all these protons in detectors that use scintillator-based tracking detectors because the minimum momenta they can measure are higher than most proton momenta.

A new neutrino-nucleus cross-section measurement is performed using the NINJA detector based on emulsion detectors to study the interactions between neutrinos and nuclei for energies ranging from hundreds of MeV to several GeV.
Nuclear emulsion is suitable for performing high precision measurements of the positions and angles of charged particles emitted from neutrino interactions since it provides sub-$\mu$m spatial resolution.
Since 2014, a series of pilot experiments~\cite{NINJA_Run4_ECC_2017,NINJA_Run4_Shifter_2017,NINJA_Run8_2020} has been run using the emulsion-based detectors.
The emulsion detector is capable of detecting slow protons with momenta as low as 200\,MeV/$c$, representing a distinct advantage compared to other detectors with higher proton momentum thresholds of approximately 400--700\,MeV/$c$~\cite{MicroBooNE_multiplicity_2019,T2K_proton_2018,MINERvA_CCQE_2013}.
This paper reports a measurement of the flux-averaged cross section of $\nu_{\mu}$ charged\nobreakdash-current interaction using an emulsion detector combined with a 65-kg iron target.
This measurement is a basis for the detailed study of charged particles produced in neutrino-nucleus interactions using the emulsion detector.

\section{Detector configuration and data samples}\label{sec:DETECTOR CONFIGURATION AND DATA SAMPLES}
\begin{figure}[htbp]
	\begin{center}
			\includegraphics[width=12cm]{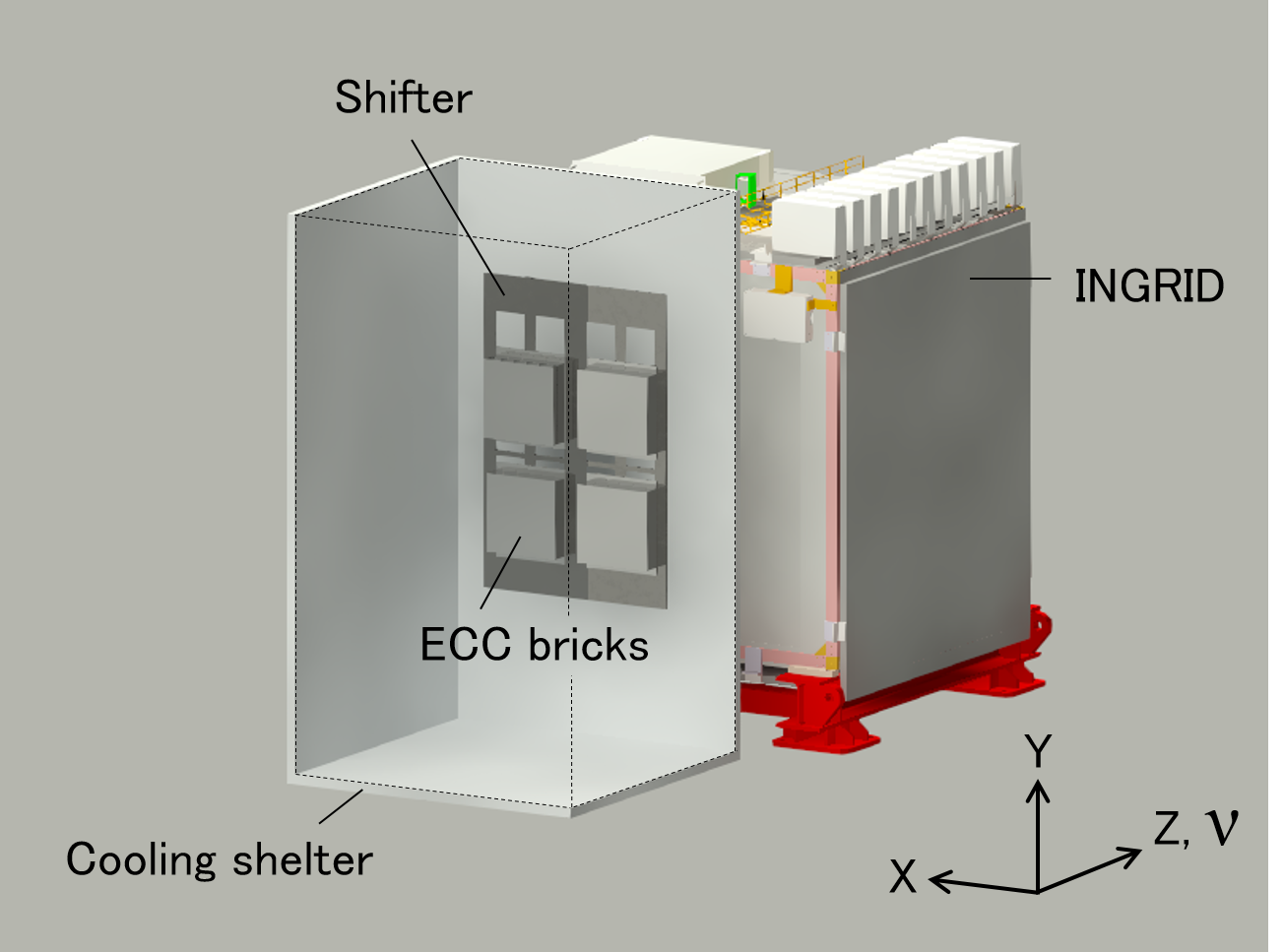}
			\caption{Schematic view of the detector. The ECC bricks and the Shifter are enclosed in a cooling shelter, which is placed in front of an INGRID module. Neutrino interactions occur in the ECC brick, and induced muons ought to be measured by the Shifter and INGRID.}
			\label{fig:detector_overallview}
	\end{center}
\end{figure}

The detector is located in the near detector hall of the T2K experiment at J\nobreakdash-PARC.
Figure~\ref{fig:detector_overallview} shows a schematic view of the detector.
The detector is a hybrid apparatus composed of an iron target emulsion cloud chamber~(ECC), an emulsion multi-stage shifter~(Shifter)~\cite{S.Takahashi_2010,S.Takahashi_2016,mizutani_2019}, and interactive neutrino grid~(INGRID) detector~\cite{INGRID_2010}.
The ECC consists of 12 basic units called bricks, which are made of emulsion films interleaved with iron plates.
The ECC bricks and the Shifter are installed upstream of the INGRID module, which is one of the near detectors in the T2K experiment.
They are placed in the order of the ECC bricks, the Shifter and the INGRID module from the beam upstream side.
The Shifter adds timing information to each track observed in an ECC brick, which helps to match the tracks with a corresponding muon track in INGRID.
The ECC bricks and the Shifter are enclosed in a cooling shelter to maintain a temperature of approximately 10\,$^{\circ}$C and protect the emulsion films from sensitivity degradation and fading.
In this study, INGRID is used as a muon range detector, from which $\nu_{\mu}$ charged\nobreakdash-current interactions are selected.
In the following, the X- and Y-axes are defined as the horizontal and vertical directions  perpendicular to the beam direction~(Z-axis), respectively.

\subsection{J\nobreakdash-PARC neutrino beamline}\label{subsec:beamline}
The J\nobreakdash-PARC accelerator provides a high-intensity 30\,GeV proton beam.
Each proton beam spill consists of eight bunches.
The width of each bunch is approximately 58\,ns, while the interval between the bunches is approximately 581\,ns.
These spills are delivered to a graphite target every 2.48\,s.
Hadrons, mainly pions, are produced by the interaction of protons with the target.
The charged pions are parallel focused by three magnetic horns.
During their flight in the decay volume, they decay primarily into muons and muon-neutrinos.
By changing the polarity of the magnetic horns, the neutrino and anti-neutrino beam modes can be switched.
Thus, an almost pure $\nu_{\mu}$ beam is delivered to the neutrino near detector hall.
The neutrino beam has energies ranging from hundreds of MeV to a few GeV at the detector location, with a peak at approximately 1\,GeV.
Further details of the J\nobreakdash-PARC neutrino beamline can be found in Ref.~\cite{Flux_JNUBEAM_2013}.

\subsection{INGRID}\label{subsec:INGRID}
\begin{figure}[htbp]
\begin{center}
	\subfigure{
		\includegraphics[clip, width=0.5\columnwidth]{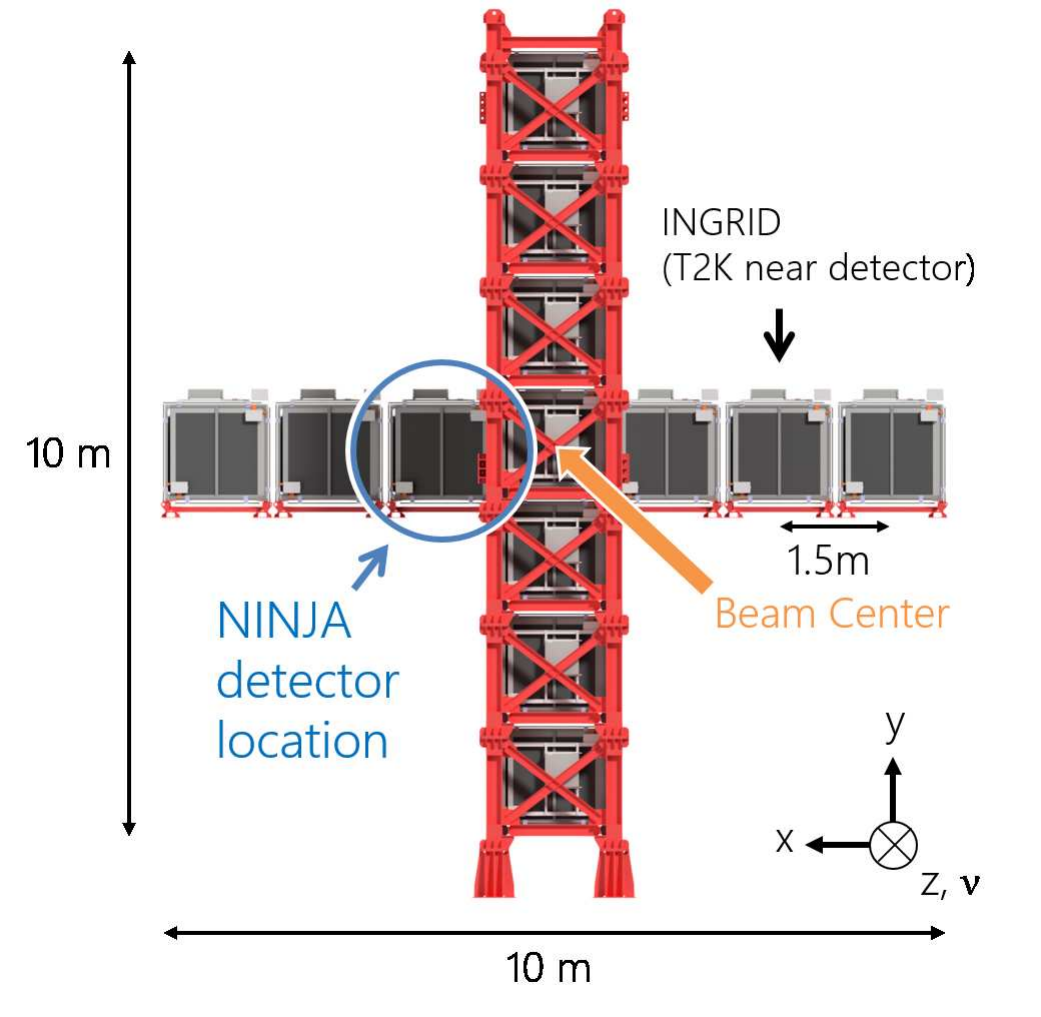}}\\
	\subfigure{
		\includegraphics[clip, width=0.6\columnwidth]{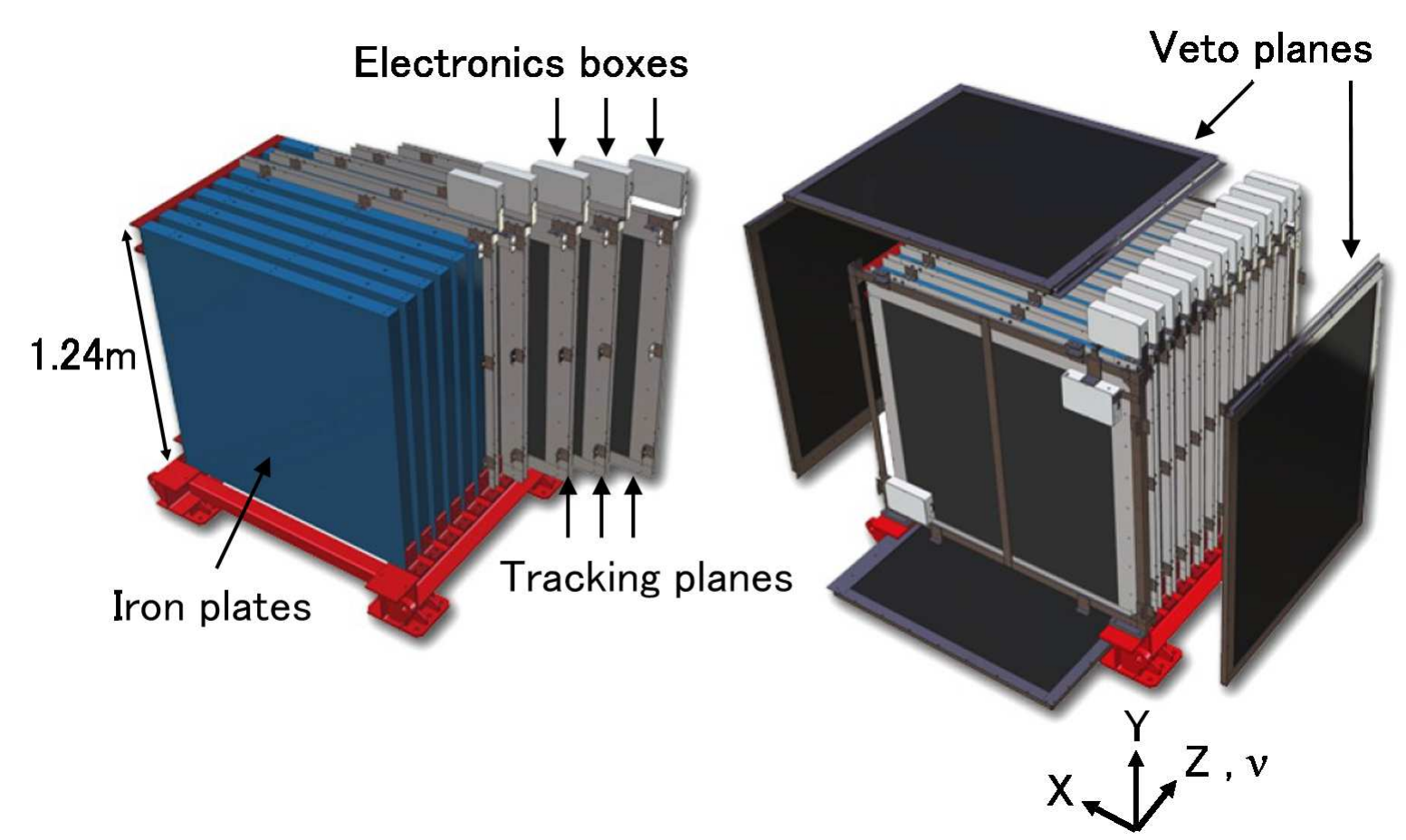}}
	\caption{(Top)~Projected view of the INGRID modules and (bottom)~exploded view of an INGRID module.}
	\label{fig:detector_ingrid}
\end{center}
\end{figure}

INGRID is the on-axis near detector for the T2K experiment located at 280\,m downstream from the proton target~\cite{T2K_2011,INGRID_2010}.
Figure~\ref{fig:detector_ingrid} top shows the position of the 14 INGRID modules arranged in a cross shape.
One of the horizontal modules next to the central module serves as a muon range detector in this measurement.
Each INGRID module comprises 11 scintillator planes interleaved with nine iron plates, as shown in the bottom half of Fig.~\ref{fig:detector_ingrid}.
Each iron plate measures 124\,cm $\times$ 124\,cm $\times$ 6.5\,cm.
Each of the 11 scintillator planes features 24 $\times$ 2 plastic scintillator bars, with alternated X- and Y- directions.
This structure makes it possible to reconstruct three-dimensional muon tracks.
The dimensions of each scintillator bar are 5\,cm $\times$ 1\,cm $\times$ 120\,cm.
The scintillation light is collected by a wavelength-shifting fiber, which is inserted into a hole in the center of the scintillator strip.
One end of the fiber is attached to a multi-pixel photon counter~(MPPC) with an optical connector for photometric measurement.

\subsection{ECC}\label{subsec:ECC}
\begin{figure}[htbp]
	\begin{center}
			\includegraphics[width=12cm]{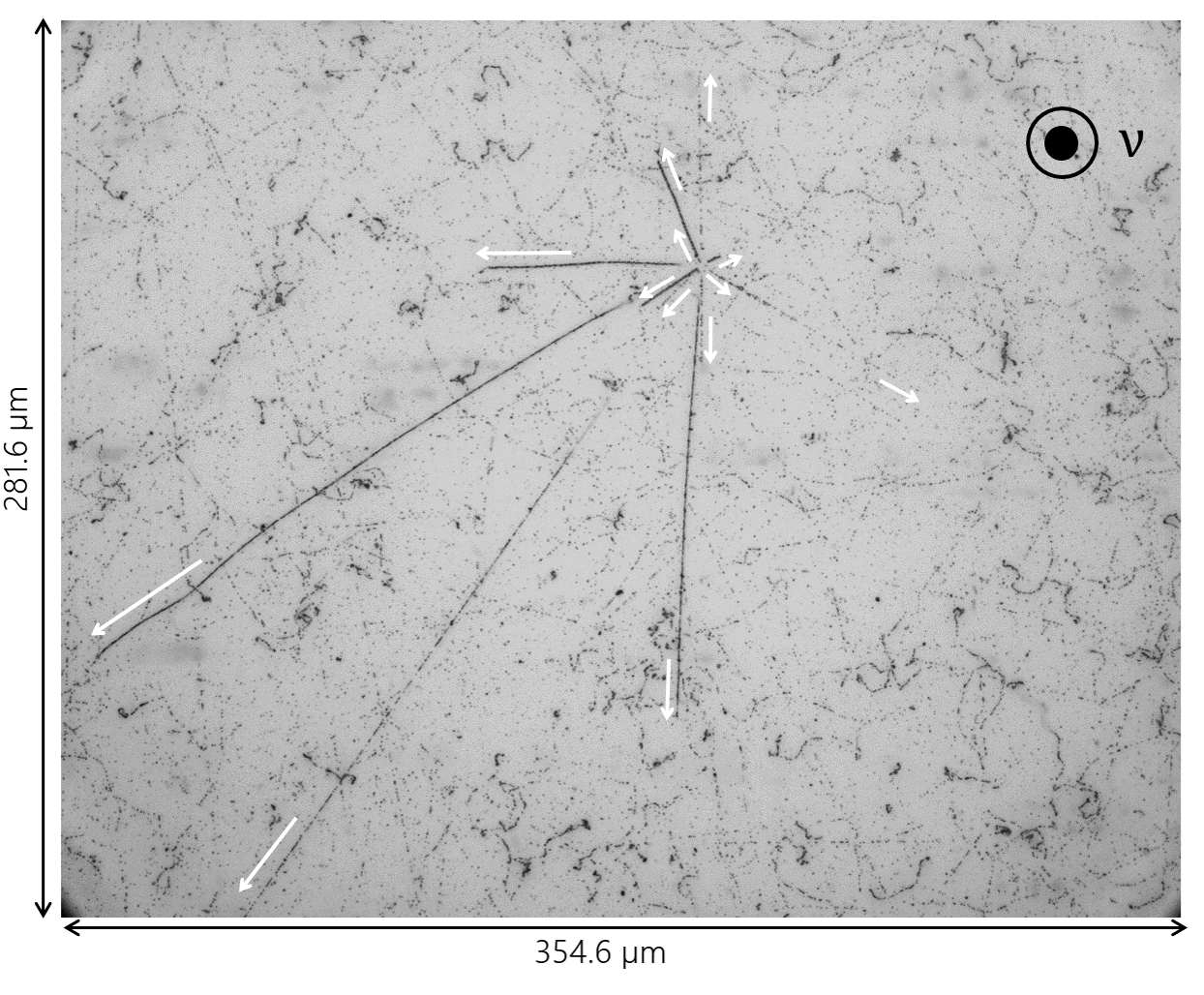}
			\caption{FTS image of charged particles emitted from a neutrino-iron interaction. Shown is an emulsion film used in this measurement. The black lines are the charged particle tracks, with the white arrows showing the direction of each track.}
			\label{fig:microscope_emulsion_track}
	\end{center}
\end{figure}

The nuclear emulsion comprises AgBr crystals embedded in gelatin.
The crystal volume occupancy of the emulsion used in this experiment is 45\%.
Properties of this emulsion are described in Refs.~\cite{morimoto_2020,nishio_2020}.
The area and thickness of the emulsion film are 25\,cm $\times$ 25\,cm and 300\,$\mu$m, respectively.
Each emulsion film is a 180\,$\mu$m polystyrene sheet with a 60-$\mu$m thick nuclear emulsion layer on each face.
Charged particle trajectories are leaving latent images in the emulsion transformed into visible rows of grains during development.
The rows of grains are measured using an optical microscope.
Figure~\ref{fig:microscope_emulsion_track} shows an image of charged particles emitted from a neutrino-iron interaction in an emulsion layer, which was acquired using a microscope system called fine track selector~(FTS)~\cite{FTS_2013,FTS_2014}.
The black lines in Fig.~\ref{fig:microscope_emulsion_track} represent the charged particle tracks.

\begin{figure}[htbp]
	\begin{center}
			\includegraphics[width=8cm]{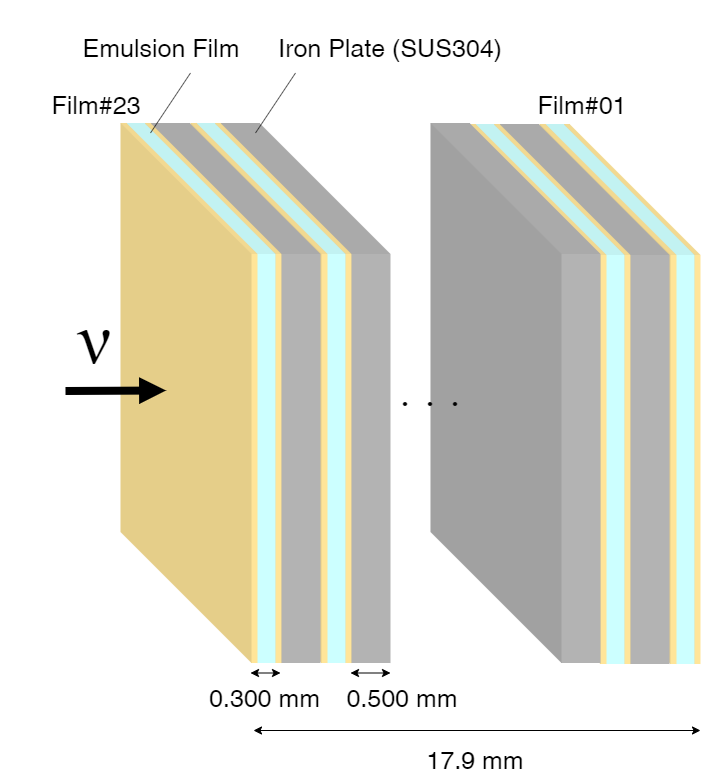}
			\caption{Structure of the iron ECC brick. Each ECC brick consists of 23 emulsion films interleaved with 22 iron plates which are made of stainless steel~(SUS304).}
			\label{fig:ECC_structure}
	\end{center}
\end{figure}

As shown in Fig.~\ref{fig:ECC_structure}, the ECC brick is composed of 23 emulsion films interleaved with 22 iron plates, each of which measures 25\,cm $\times$ 25\,cm $\times$ 0.05\,cm.
The iron plates are made of stainless steel~(SUS304) instead of pure iron in order to avoid chemical reactions in contact with emulsion films.
It consists of iron~(72.3\%), chromium~(18.1\%), nickel~(8.0\%), and other contaminations, including manganese, silicon, phosphorus, and sulfur~(1.6\% in total).
Numbers of neutrons and protons in nuclei composing the stainless steel are close to those in iron.
Furthermore, the neutron to proton number ratio in the stainless steel is 1.149, compared to 1.150 for iron.
Therefore, in the following analysis, the target material in the ECC bricks is treated as iron.
For this pilot experiment, the ECC comprised 12 bricks, with the iron plates having a total mass of 65\,kg.

\begin{figure}[htbp]
	\begin{center}
			\includegraphics[width=16cm]{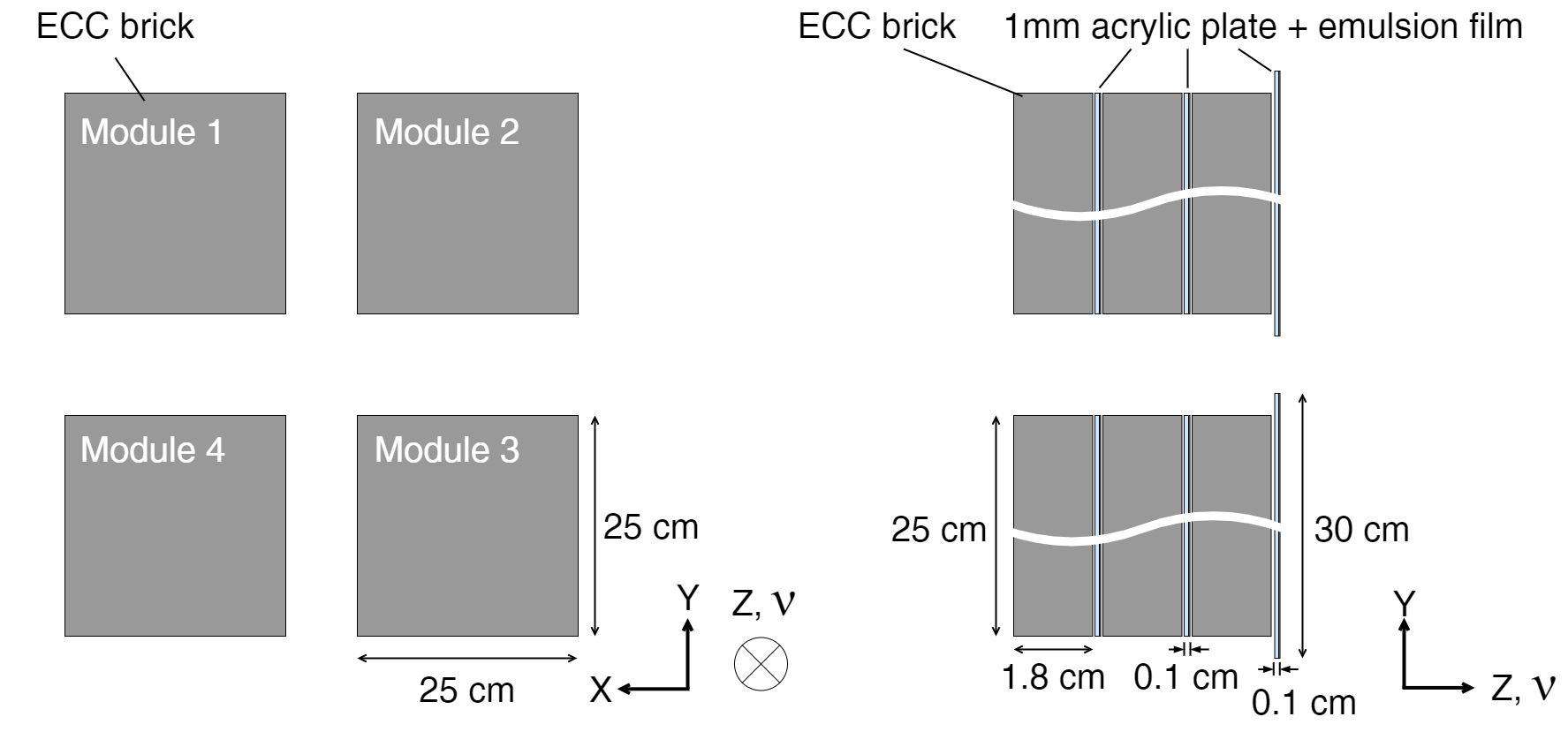}
			\caption{(Left) Front view and (right) side view of the ECC bricks. Four ECC bricks were placed in the XY-plane, with this configuration repeated to produce a structure that was three bricks deep along the Z-direction.}
			\label{fig:ECC_XY_YZ_plane}
	\end{center}
\end{figure}

As shown in Fig.~\ref{fig:ECC_XY_YZ_plane}, four ECC bricks were placed in the XY-plane in a square configuration, with another two bricks placed behind them along the Z-direction, i.e., the four-brick square was three bricks deep.
In addition, a subsidiary emulsion film was placed between the ECC bricks as well as on the brick face farthest downstream to facilitate track connection, as shown in the right-hand schematic of Fig.~\ref{fig:ECC_XY_YZ_plane}.
One of the bricks was taken out from the detector about one month after the beginning of the exposure and developed to check the emulsion quality.
This brick was not used in the following analysis.

\subsection{Shifter}\label{subsec:Shifter}
\begin{figure}[htbp]
	\begin{center}
			\includegraphics[width=12cm]{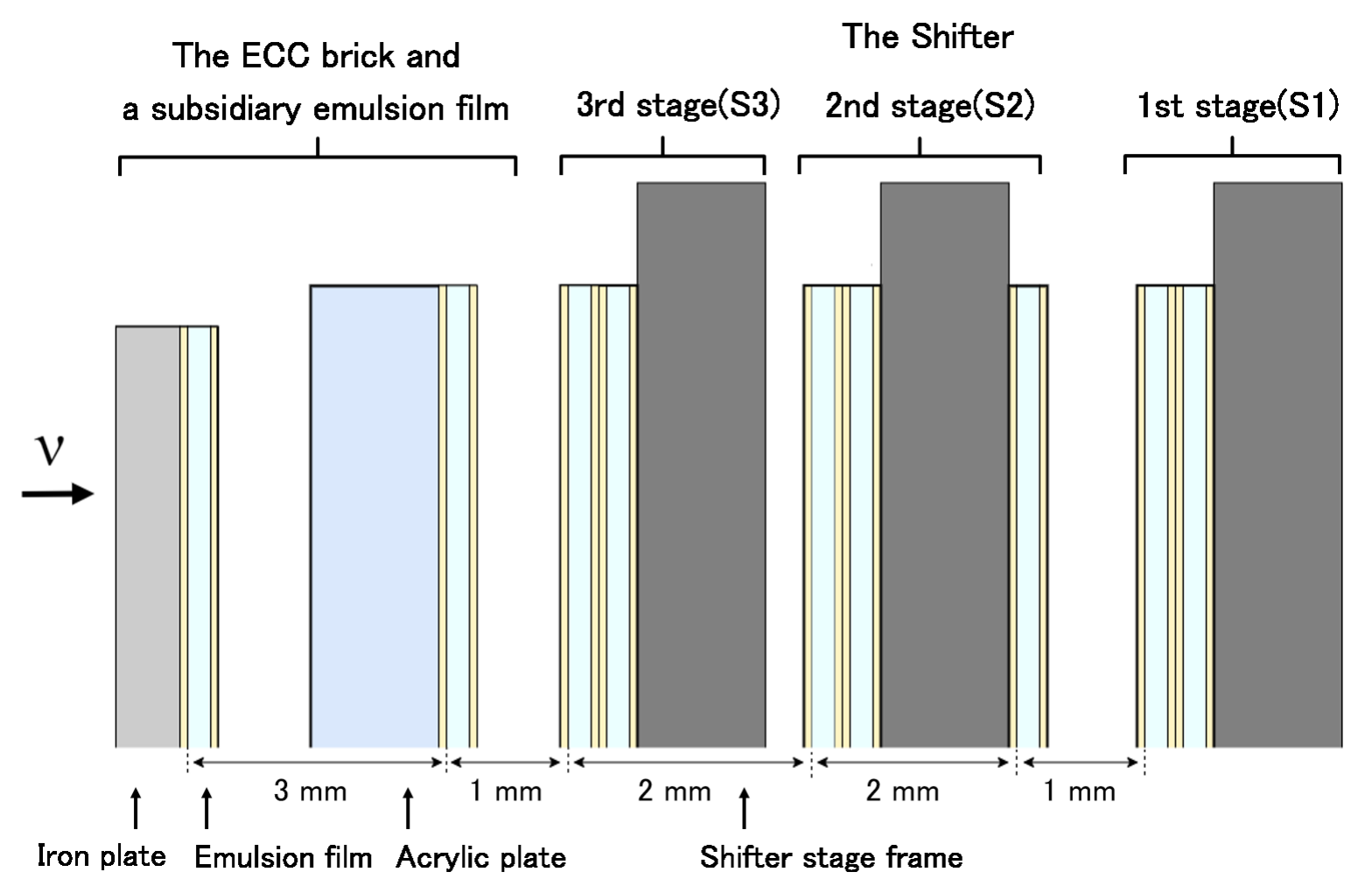}
			\caption{Elevational structure of the Shifter. Two emulsion films are mounted on both S1 and S3, with three mounted on S2. }
			\label{fig:shifter_structure}
	\end{center}
\end{figure}

\begin{figure}[htbp]
	\begin{center}
			\includegraphics[width=10cm]{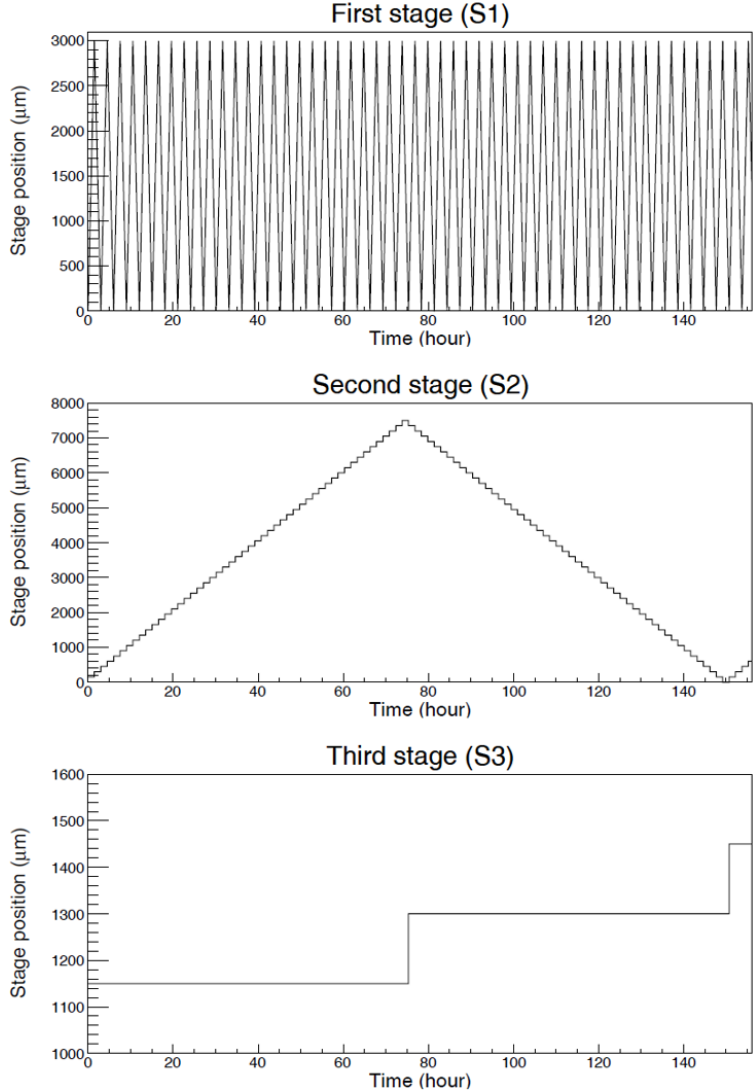}
			\caption{Respective positions of each Shifter stage over time. Each stage is driven at a different speed to add timing information to the ECC tracks.}
			\label{fig:stage_position_0_156}
	\end{center}
\end{figure}

The emulsion shifter technique was originally developed for a balloon experiment to study cosmic-ray electrons~\cite{K.Kodama_2006} and was introduced into the GRAINE experiment~\cite{S.Takahashi_2016,S.Takahashi_2018}.
The Shifter is composed of three stages~(S1, S2, S3), as shown in Fig.~\ref{fig:shifter_structure}.
Seven films, each with an area of 25\,cm $\times$ 30\,cm, are mounted on the three stages in a 2:3:2 ratio.
Stages S1 and S2 are separated by a gap of 3\,mm, while S2 and S3 are separated by a gap of 2\,mm.
Each stage is driven at a different speed along the Y-direction in order to add timing information to ECC tracks.
Figure~\ref{fig:stage_position_0_156} shows the respective positions of each stage over time: S1 has a cyclical motion pattern, moving at a speed of 0.553\,$\rm \mu$m/s and a stroke of 3000\,$\rm \mu$m; S2 shifts at each instant when S1 changes direction, and is driven by a stepping motor with a step size of 150\,$\rm \mu$m and a stroke of 7500\,$\rm \mu$m~(it has a repetition time of 1.51\,h); S3 shifts when S2 changes direction, and is driven by a stepping motor with a step of 150\,$\rm \mu$m~(it has a repetition time of 3.1 days).
A full cycle of the Shifter operation lasts about 155 days.
The Shifter allows the addition of time information to the ECC tracks up to 155 days.

\subsection{Data samples}\label{subsec:data_sampls}
The ECC bricks and the Shifter were exposed to the neutrino beam between February and May 2016.
The neutrino beamline was operated in both neutrino and anti-neutrino beam modes.
There were two periods of exposure in the neutrino beam mode: the first was from February 1--3, and the second was from May 19--27.
After live-time correction of the detectors, we analyzed data samples in the neutrino beam mode, corresponding to $4.0 \times 10^{19}$ protons on target~(POT).

\section{Monte Carlo simulation}\label{sec:MC}
The signal and background events, neutrino flux, and detection efficiency were estimated using Monte Carlo~(MC) simulations.
These MC simulations consist of three parts: (i) JNUBEAM~\cite{Flux_JNUBEAM_2013} to predict the neutrino flux, (ii) NEUT~\cite{NEUT_2002,NEUT_2009} to model the interactions between neutrinos and nuclei, and (iii) a GEANT4~\cite{Geant4_2003,Geant4_2006,Geant4_2016}-based framework to simulate the detector response.
The MC predictions were normalized with respect to the POT value and the target mass.

\subsection{Neutrino beam}\label{subsec:MC_flux}
\begin{figure}[htbp]
	\begin{center}
		\includegraphics[width=10.5cm]{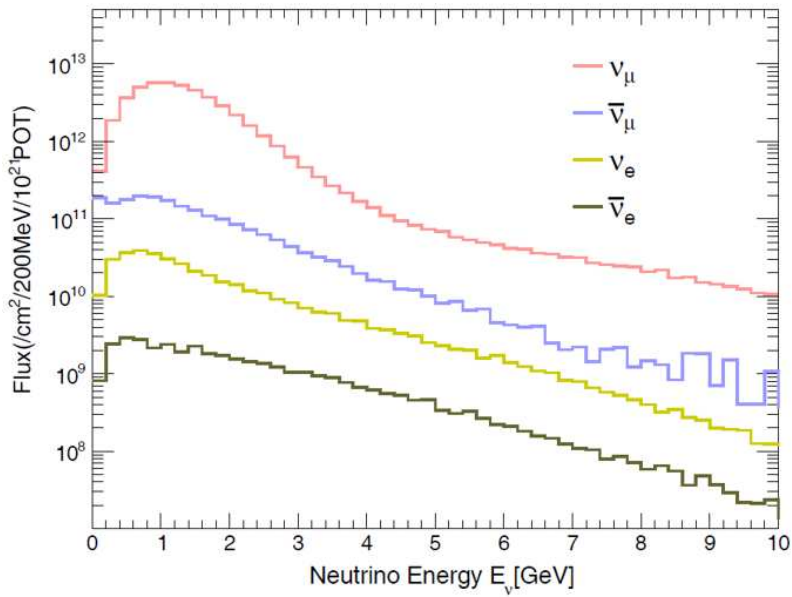}
		\caption{Neutrino energy spectrum of each beam component at the NINJA detector in the neutrino beam mode. These spectra were predicted using JNUBEAM.}
		\label{fig:NINJA_flux}
	\end{center}
\end{figure}

The neutrino flux at the NINJA detector is estimated using JNUBEAM, which was developed to predict the flux and spectrum of neutrinos at the T2K detectors and is based on a GEANT3 framework~\cite{Geant_1994}.
We used JNUBEAM version 13av6.1.
FLUKA2011.2~\cite{FLUKA_2005,FLUKA_2014} was used to simulate the hadronic interactions of primary protons on the graphite target.
Then, JNUBEAM takes the secondary particle information simulated using FLUKA, and models their propagation, interaction, and decay events.
The hadronic interaction simulation was tuned using hadron production data from experiments such as CERN NA61/SHINE ~\cite{NA61_2014}, where a combination of measurements with a replica of the T2K target~\cite{NA61_t2k_replica_target_2013,NA61_t2k_replica_target_2016} and a thin graphite target~\cite{NA61_t2k_thin_target_2011,NA61_t2k_thin_target_2012,NA61_t2k_thin_target_2014,NA61_t2k_thin_target_2016} was used.
In the neutrino beam mode, the mean energy and fraction of $\nu_{\mu}$ components are 1.49\,GeV and 94.9\%.
The fraction of $\bar{\nu}_{\mu}$ components is 4.3\%, with $\nu_{e}$ and $\bar{\nu}_{e}$ components representing the remaining 0.8\%.
Figure~\ref{fig:NINJA_flux} shows the neutrino energy spectrum of each beam component at the NINJA detector in the neutrino beam mode. 

\subsection{Event generation}\label{subsec:MC_interaction}
\begin{figure}[htbp]
	\begin{center}
		\includegraphics[width=10cm]{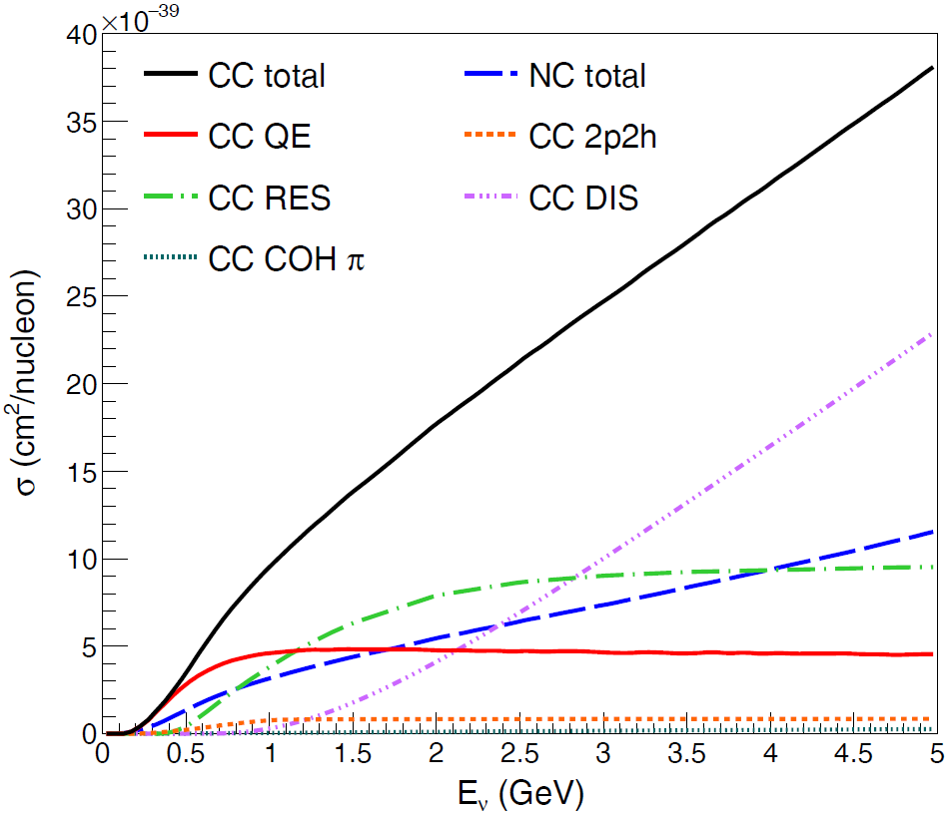}
		\caption{Neutrino-nucleus cross sections per nucleon of an iron nucleus predicted by NEUT.}
		\label{fig:Xsec_curve_numu}
	\end{center}
\end{figure}

The Super-Kamiokande~\cite{SK_2003} and T2K experiments use the NEUT neutrino event generator.
In addition to simulating primary neutrino interactions, NEUT also simulates final state interactions~(FSIs), such as scattering, absorption, particle production, and charge-exchange of hadrons produced by neutrino interactions in the nuclear medium prior to escape.
We used version 5.4.0 of NEUT.
To predict the signal and background events in the ECC bricks, $\nu_{\mu}$, $\bar{\nu}_{\mu}$, $\nu_{e}$, and $\bar{\nu}_{e}$ interactions on iron were generated using NEUT.
The MC prediction for an iron target was adapted to the stainless steel target using the difference in the fractions of protons and neutrons between iron and the stainless steel.
In addition, neutrino interactions in the upstream wall of the detector hall and the INGRID modules were generated as background sources.
The neutrino interaction models and nominal parameters used in NEUT are listed in Table~\ref{tab:NEUT_model_parameter}.
Charged\nobreakdash-current~(CC) quasi-elastic~(QE) and neutral-current~(NC) elastic scatterings, two-particle-two-hole~(2p2h) excitations, CC and NC resonant interactions~(RES), coherent pion productions~(COH\,$\pi$), and deep inelastic scatterings~(DIS) were simulated.
The one-particle-one-hole~(1p1h) model by Nieves $\textit{et al.}$~\cite{1p1h_nieves_2012,rpa_nieves_2012} was used to simulate the CCQE.
In this model, a local Fermi gas~(LFG) model with random phase approximation~(RPA) corrections is used for the nuclear model, and the axial mass $M_{\rm A}^{\rm QE}$ is set to 1.05~GeV/$c^{2}$.
Nieves $\textit{et al.}$ have also modeled the 2p2h interaction~\cite{2p2h_nieves_2011}.
The RES was simulated using the Rein-Sehgal model~\cite{res_rein_sehgal_1981}, and the axial mass $M_{\rm A}^{\rm RES}$ is set to 0.95~GeV/$c^{2}$.
In addition, we used the COH\,$\pi$ model described by Rein-Sehgal model in Refs.~\cite{coh_rein_sehgal_1983,coh_rein_sehgal_2007}.
To describe DIS, we applied parton distribution function~(PDF) GRV98 with Bodek and Yang correction~\cite{dis_pdf_1998,dis_pdf_2003,dis_pdf_2005}.
NEUT models the FSI for hadrons using a semi-classical intra-nuclear cascade model~\cite{NEUT_2009,pion_fsi_2014,pion_fsi_2019}.
Figure~\ref{fig:Xsec_curve_numu} shows the neutrino-nucleus cross sections per nucleon of an iron nucleus predicted by NEUT.
\begin{table}[htbp]
	\caption{Neutrino interaction models used in the nominal MC simulation.}
	\label{tab:NEUT_model_parameter}
	\begin{center}
	\begin{tabular}{ll}
		\hline
		\hline
		 Interaction & Model \\
		\hline
		 CCQE & 1p1h model by Nieves $\textit{et al.}$~\cite{1p1h_nieves_2012,rpa_nieves_2012} \\
		 & LFG with RPA correction~(${\textit{M}}^{\textrm{QE}}_{\textrm{A}}$=1.05\,GeV/$c^{2}$) \\
		\hline
		 2p2h & 2p2h model by Nieves $\textit{et al.}$~\cite{2p2h_nieves_2011} \\
		\hline
		 RES & Model described by Rein-Sehgal~\cite{res_rein_sehgal_1981}~(${\textit{M}}^{\textrm{RES}}_{\textrm{A}}$=0.95\,GeV/$c^{2}$) \\
		\hline
		 COH\,$\pi$ & Model described by Rein-Sehgal~\cite{coh_rein_sehgal_1983,coh_rein_sehgal_2007} \\
		\hline
		 DIS & GRV98 PDF with Bodek and Yang correction~\cite{dis_pdf_1998,dis_pdf_2003,dis_pdf_2005} \\
		\hline
		 FSI & Semi-classical intra-nuclear cascade model~\cite{NEUT_2009,pion_fsi_2014,pion_fsi_2019} \\
		\hline
		\hline
	\end{tabular}
	\end{center}
\end{table}

\subsection{Detector response}\label{subsec:MC_detector}
The detector simulation for the particles was developed with a GEANT4 framework.
We used GEANT4 version 9.2.1 and the QGSP~BERT physics list~\cite{QGSPBERT_2009}.
The geometry of the ECC bricks, Shifter, INGRID modules, and wall of the detector hall were modeled for the detector simulation.
The base track~(described in Sec.~\ref{subsec:track_recon_ecc}) was reconstructed using the positions that charged particles pass through on both faces of the polystyrene sheet.
Therefore, the detection efficiency of the base tracks evaluated by the data was incorporated in the MC simulation.
A volume pulse height~(VPH)~\cite{VPH_2006} corresponding to an energy deposit in the emulsion film was reconstructed on the basis of the correlation between a given slope, momentum, and VPH in the data.
This process was repeated for the MC simulation of the track connections between the films in individual ECC bricks, between the ECC bricks, and between the ECC bricks and INGRID.
For the track connections between the ECC bricks and the Shifter, we used the connection efficiency based on the data.
The background events produced by cosmic rays and misconnected events between the ECC bricks, the Shifter, and INGRID were estimated using the data rather than the MC simulation.

\section{Track reconstruction}\label{sec:RECONSTRUCTION}
\subsection{Track reconstruction in the ECC bricks}\label{subsec:track_recon_ecc}
The track pieces recorded in an emulsion layer are called ``micro tracks.''
The positions $(x, y)$ and slopes~(tan$\theta_{x}$, tan$\theta_{y}$) of the micro tracks were measured using the hyper-track selector~(HTS)~\cite{HTS_2017}.
The HTS recognizes a series of grains on a straight line as a micro track by taking 16 tomographic images in the emulsion layer.
The slope acceptance of the HTS was set to $|$tan$\theta_{x(y)}$$|$$<$1.7~($|\theta_{x(y)}|\lesssim 60^{\circ}$).
The track angle $\theta$ is defined as the angle with respect to the Z-direction.
The pulse height~(PH)~\cite{PH_2004} and VPH of the track were measured.
The PH is defined as the number of tomographic images that have pixels associated with the track, while the VPH is the total number of pixels  associated with the track in all 16 tomographic images.
The PH and VPH measure the mean energy loss of the charged particles.
A single scan covers an area of 130\,mm $\times$ 90\,mm, with each emulsion film covered by six scans.
After scanning, the tracks are reconstructed via a NETSCAN~\cite{DOUNUT_2002,NETSCAN2_2012}-based procedure.
The tracks connecting the positions of micro tracks on both sides of the polystyrene sheet are called ``base tracks,'' and are used as track segments in this analysis.
The base track detection efficiency ranges between 95 and 99\%, with variation caused by individual film differences.
After reconstructing the base tracks, the rotation, slant, parallel translation, and gap between emulsion films are adjusted.
This alignment process determines the relative positions of the films during the beam exposure.
Following the alignment process, the ECC tracks are reconstructed by connecting the base tracks both in adjacent films and in films separated by one or two films.
The slope- and position-related tolerances associated with the base track connections are defined as functions of the track slope.
The connection efficiency exceeds 99\%.
In this analysis, the ECC tracks are required to pass through at least one iron plate and two emulsion films.
The momentum threshold for proton tracks is around 200\,MeV/$c$, while that for charged pion tracks is around 50\,MeV/$c$.

\subsection{Time-stamping to the ECC tracks}\label{subsec:timestamp}
As shown in Sec.~\ref{subsec:Shifter}, the Shifter comprises seven emulsion films mounted on three stages.
The scanning area and the slope acceptance of the films in the Shifter are equivalent to those of the ECC brick films.
Each Shifter film is covered by eight scanning areas.
The Shifter tracks were also reconstructed using the NETSCAN software package.
The tracks in the ECC bricks and the Shifter are connected in order to add timing information to the ECC tracks.
Prior to the connecting procedure, the ECC bricks are aligned with each of the Shifter stages using as reference the fixed position of the stages for one week after the Shifter has completed its operation.
After this alignment, the tracks between the ECC bricks and each Shifter stage are connected in the following order: first, the tracks between the most downstream film of the ECC brick and S3 are connected; then, the tracks between S3 and S2 are connected; finally, the tracks between S2 and S1 are connected.
%
%
%
%
The tracks are connected using slope- and position-based matching, with slope- and position-related tolerances set to 0.025 and 75\,$\mu$m in the XZ- and YZ-planes, respectively.

\subsection{Track reconstruction in INGRID}\label{subsec:track_recon_ingrid}
The track reconstruction and selection processes used in this analysis for the INGRID detector are similar to those used in the T2K experiment~\cite{INGRID_CCinclusive_Xsec_2014,INGRID_CCinclusive_Xsec_2016}.
Each track is composed of a series of hits, where a hit is defined as an MPPC channel that exceeds a signal equivalent to 2.5 photoelectrons.
The hits are clustered within $\pm$ 50\,ns from the average hit time.
A tracking plane that contains at least one hit in both X-~(horizontal) and Y-~(vertical) layers is defined as an active plane.
Events are required to have at least three active planes, corresponding to a muon momentum threshold of approximately 300\,MeV/$c$.
Two-dimensional tracks in the XZ- and YZ-planes are reconstructed independently using a track reconstruction algorithm based on cellular automaton~\cite{cellular_automaton_2005}, while three-dimensional tracks are reconstructed by merging track pairs in the XZ- and YZ-planes.
Events are selected within $\pm$ 100\,ns from the event timing, which is defined as the hit timing of the channel with the largest number of photoelectrons.
In this analysis, the INGRID tracks are required to start at the most upstream plane.

\subsection{Track matching}\label{subsec:track_matching}
\begin{figure}[htbp]
	\begin{center}
		\includegraphics[width=10cm]{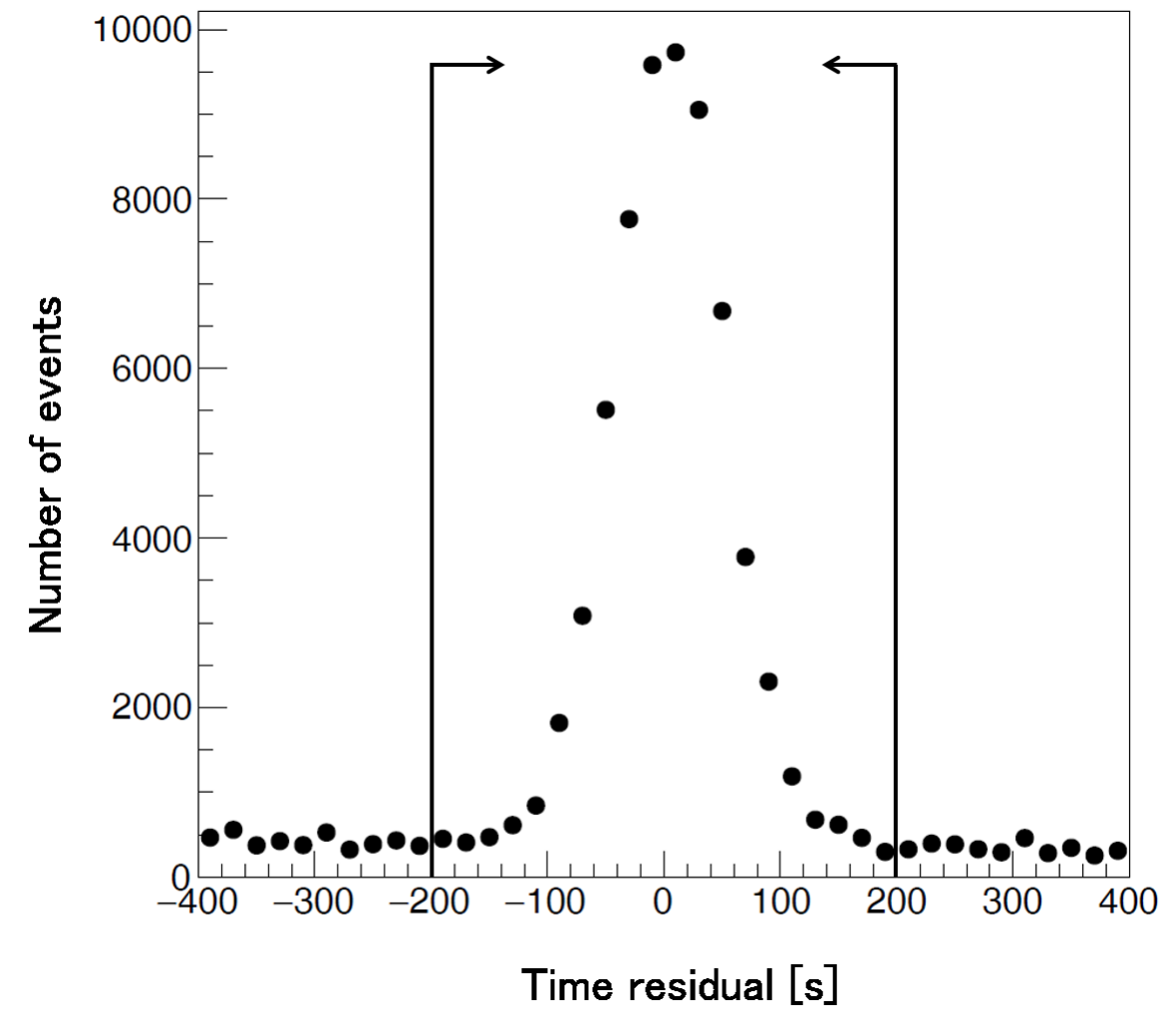}
		\caption{Time residuals between the ECC events and INGRID events. The data samples are all events that occurred during the Shifter operation. The standard deviation of the time residuals distribution (approximately 50\,s) corresponds to the estimated time resolution of the Shifter. The black lines and arrows represent the time tolerance for the ECC--INGRID track matching.}
		\label{fig:time_residual}
	\end{center}
\end{figure}

\begin{figure}[htbp]
	\begin{center}
		\includegraphics[width=10cm]{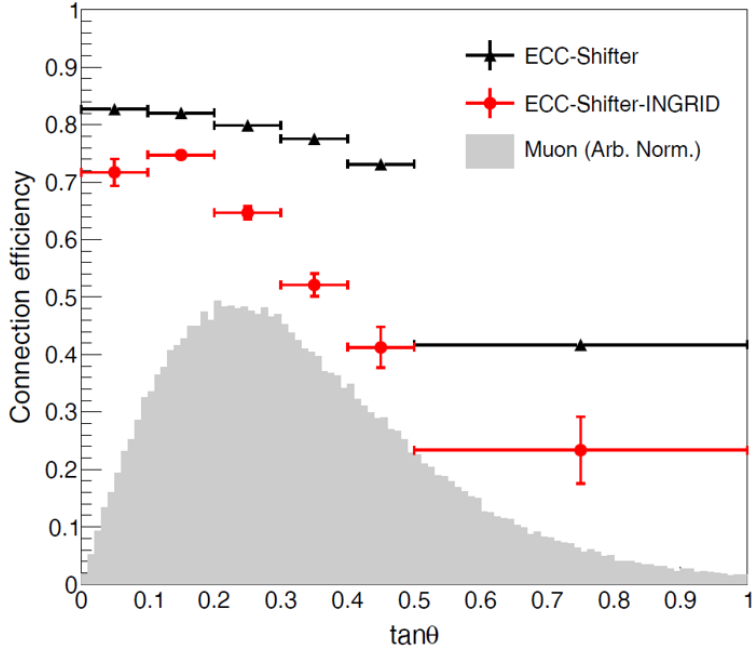}
		\caption{Connection efficiencies of the muon tracks among the ECC bricks, the Shifter, and INGRID as a function of tan$\theta$. The black and red markers represent the connection efficiencies, and their vertical errors represent statistical errors. The gray histogram represents with an arbitrary normalization the expected tan$\theta$ distribution of muons emitted from $\nu_{\mu}$ CC interactions in the ECC fiducial volume described in Sec.~\ref{sec:EVENT_SELECTION}.}
		\label{fig:connection_efficiency}
	\end{center}
\end{figure}

After connecting the tracks between the ECC bricks and the Shifter, track matching between the ECC bricks and the INGRID module is performed.
The ECC tracks are extrapolated to the most upstream plane of INGRID and matched with an INGRID track using the slope, position, and timing information.
Each ECC event is required to have a time residual within 200\,s from the INGRID event timing.
The slope- and position-related tolerances are set to 0.100 and 5.0\,cm in the XZ- and YZ-planes, respectively.
Figure~\ref{fig:time_residual} shows the time residuals between the ECC and INGRID events.
The standard deviation of the time residuals distribution represents the Shifter time resolution, which was approximately 50\,s in this study.
The connection efficiencies of muon tracks among the ECC bricks, the Shifter, and INGRID are shown in Fig.~\ref{fig:connection_efficiency}.
The connection inefficiencies are caused by the following reasons:
multiple scattering of muons;
muons stopped before penetrating three layers of INGRID;
angle acceptance of INGRID.
Events with more than one possible connections, representing approximately 4\% of the total, were not used.

\section{Event reconstruction and the selection of $\nu_{\mu}$ CC interactions}\label{sec:EVENT_SELECTION}
In this analysis, the $\nu_{\mu}$ CC interactions in the iron target are defined as signal events.
Each step of the event selection process is described below.

\begin{figure}[htbp]
\begin{center}
	\subfigure{
		\includegraphics[clip, width=0.4\columnwidth]{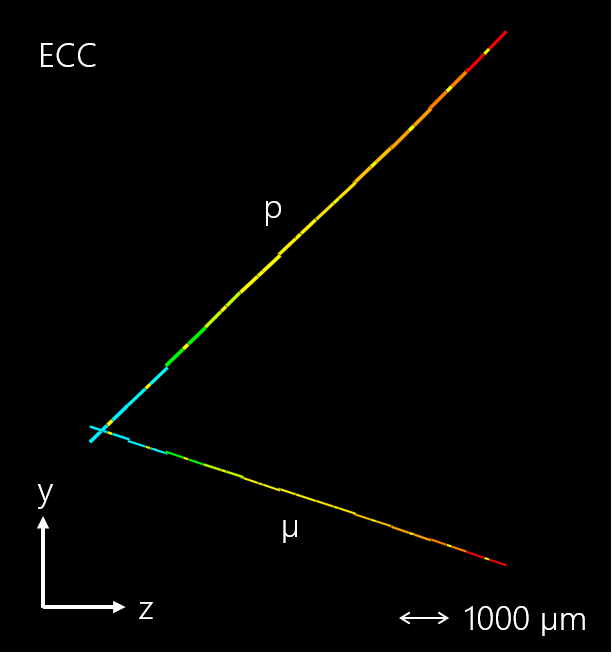}}
	\subfigure{
		\includegraphics[clip, width=0.55\columnwidth]{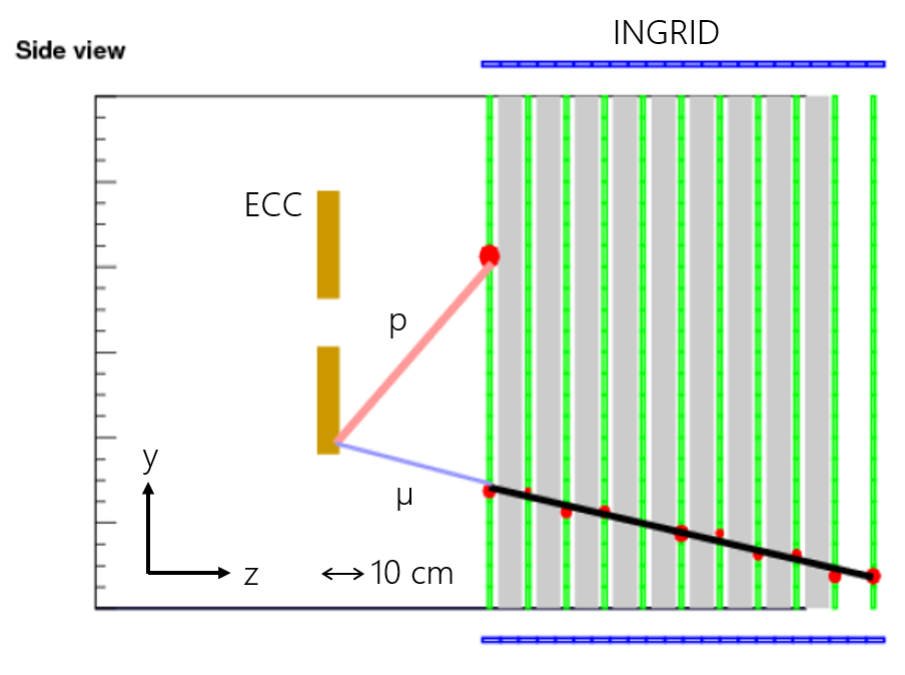}}
	\caption{Event display of a neutrino-iron CC interaction candidate.  The left-hand side of the figure shows the event display in the ECC brick, while the right-hand side shows the event display both in the ECC bricks and in INGRID. On the left-hand side, with their colors representing the track segment and their width indicating the VPH. On the right-hand side, the blue and pink lines are the ECC tracks extrapolated to INGRID, with the blue line representing a muon candidate and the pink line representing a proton-like track. The width of the blue and pink lines indicates the VPH. The red markers represent hits and their size represent deposited photoelectrons, and the black line represents the reconstructed track in INGRID.}
	\label{fig:ed_ecc_ingrid}
\end{center}
\end{figure}

\begin{enumerate}
\item ECC--Shifter--INGRID track matching\\
The ECC tracks that are matched between the ECC bricks, Shifter, and INGRID were selected as muon candidates from $\nu_{\mu}$ CC interactions.
A total of 9\,397 such events were selected.
\\

\item Scanback \\
The muon candidates were traced back from INGRID to the neutrino interaction vertices in the ECC bricks.
This procedure is known as the scanback method~\cite{DOUNUT_2002,ScanBack_1979,ScanBack_1997,ScanBack_2013,DOUNUT_2001}.
In this method, if no track satisfying slope- and position-related tolerances is found in three consecutive films, the most upstream track segment of a muon candidate is defined as its starting segment.
The iron plate on the upstream side of the starting segment is defined as the interaction plate.
\\

\item Fiducial volume cut\\
Most of the muon candidates are so-called sand muons, which are produced by neutrino interactions in the upstream wall of the detector hall.
Edges of starting segments originate from the starting positions of many sand muons and cosmic rays entering the scanning area, and the edges are located at around positions 1--2\,mm from the borders of the scanning area.
An area 5\,mm inside from the starting edges is kept as the fiducial scanning area, with the average fiducial scanning area of each film measuring 116\,mm $\times$ 78\,mm.
In the Z-axis direction, the fiducial volume~(FV) is defined as the volume between the fourth film from the upstream face and the second film from the downstream face of each ECC brick.
As a result, the target mass in the FV is 42\,kg.
The muon candidate tracks were extrapolated from the starting segment positions to the positions in the three upstream films and defined as outbound FV tracks if they are escaping from the fiducial area.
If a muon candidate started from a film that is damaged by scratches or within three films downstream of the damaged film, it was excluded from the neutrino interaction candidates in the FV.
Each film features four 3-mm diameter holes to facilitate film development.
If the muon candidates passed through holes when the tracks were extrapolated from the starting segments to the three upstream films, they were excluded from the interaction candidates in the FV.
After these cuts, there remained 236 events as interaction candidates occurring in the FV of the ECC bricks. A total of 9\,002 events were excluded as sand muon candidates, and 159 events were discarded due to the damages and holes on the films. 
\\

\item Manual microscope check\\
A process called ``manual check'' consists in a careful examination with a microscope of the region around the starting segment of the muon candidates.
First, the film just upstream of the starting segment is examined.
If a base track is found that can be connected to the starting segment, it is defined as the new starting segment.
The FV cuts are applied to this new starting segment and some additional events are rejected as possible sand muons.
The other role of the manual check is to determine in which material the neutrino interaction took place.
If the track is starting inside the emulsion layer, the interaction is considered as occurring in the emulsion.
If the track is observed only in the emulsion layer on the downstream side of the polystyrene sheet, the interaction is considered as occurring in the polystyrene.
The other events are classified as interactions in the iron target.
As a result of the manual check, 203 events were defined as interactions in the iron target, 13 events as interactions in the emulsion and 14 events as interactions in the polystyrene.
In addition, 6 events were excluded as possible sand muon tracks.
The relative rates of interactions in the different materials are consistent with their mass ratios within the statistical uncertainty.
In the MC simulations, the efficiency of the manual check is assumed to be 100\%.
\\

\item Partner track search\\
The tracks attached to the muon candidates are charged hadrons from the neutrino interactions and are called ``partner tracks.''
In this study, we defined partner tracks with VPH $<$ 150 as thin tracks, whereas tracks with VPH $\geq$ 150 were defined as black tracks.
Our search for partner tracks was performed under the following conditions: for thin tracks, the minimum distance between the muon candidate and its partner track had to be less than 50\,$\mu$m.
For black tracks, this minimum distance had to be less than 60\,$\mu$m.
In both cases, the distance along the Z-direction between the starting segment and the position of closest approach had to be less than 800\,$\mu$m.
Moreover, we required the thin tracks to have at least three track segments, while the black tracks were required to have at least two track segments.
If multiple tracks for a particular event were connected to INGRID, the track with the highest momentum~(see sub-section 7) was assumed to be the muon candidate.
For duplicate events, whereby two tracks were connected to INGRID, the higher momentum track of the two was kept as a muon candidate and the other was discarded.
\\

\item Kink event cut\\
Two track events for which the opening angle~$\alpha$ of the track pair is almost 180$^{\circ}$ typically represent background events from sand muons or cosmic rays.
If a charged particle coming from the wall exceeds the track connection tolerance, the resulting event looks like a two-track event, consisting of a forward track and a backward track connected at a vertex.
Such events are called ``kink events.''
Kink events are characterized by their large opening angle: in the region cos$\alpha$$<$$-$0.96, the background fraction is 98.2\% according to our MC study.
We conducted a particle identification process~\cite{NINJA_Run6_TAUP_2020,NINJA_Run8_2020} based on the momentum and the VPH of the partner track, which allows to separate pion-like and proton-like tracks.
Two-track events consisting of a muon candidate and a pion-like track with an opening angle in the region of cos$\alpha$$<$$-$0.96 were assumed as kink events, and were discarded.
In contrast, two-track events consisting of a muon candidate and a proton-like track were kept.
As a result of the kink cut process, 7 events were rejected.
\\

\item Momentum consistency check\\
There are two methods for estimating the momentum of a muon.
One involves measuring its multiple Coulomb scattering in the ECC bricks, while the other is to measure its energy from the track range in the ECC bricks and the INGRID detector.
The values measured by the two methods can be compared to exclude misconnected backgrounds.
Muons were considered to exhibit momentum inconsistencies if they met the following criteria: if the momentum estimated by an angular scattering measurement~\cite{MCS_angular_method_2012} was greater~(smaller) than 218\%~(17\%) of that measured by the range;
if the momentum estimated by a positional scattering measurement~\cite{MCS_angular_coordinate_method_2007} was greater~(smaller) than 352\%~(45\%) of that measured by the range.
The maximum and minimum limits were based on the two-sigma confidence interval of the momentum measurement accuracy.
In the case of muons passing through or side-escaping INGRID, only the minimum limits were considered.
The momentum consistency check led to the exclusion of 12 events.
These events can be mainly due to misconnection of the ECC track to INGRID or to cosmic muon coming from downstream and stopping in an ECC brick.
\\

\end{enumerate}
Figure~\ref{fig:ed_ecc_ingrid} shows the display of a selected neutrino-iron interaction candidate.
The event contains a muon and a proton-like track.
The number of selected events remaining after each selection step is summarized in Table~\ref{tab:num_sel_events}.
The purity is defined as the fraction of $\nu_{\mu}$ CC interactions on iron in the MC sample.
Finally, a total of 183 events were confirmed as $\nu_{\mu}$ CC interactions on the iron target, corresponding to 188.8 events in the MC sample, in good agreement within the statistical uncertainty.
The events predicted by the MC simulation were categorized as follows:
88.2\% of the events were signals produced by $\nu_{\mu}$ CC interactions,
4.8\% were misconnected backgrounds,
3.4\% were hadron interactions caused by neutrons, protons and charged pions emitted from the neutrino interactions in the upstream wall,
2.7\% were events arising from $\bar{\nu}_{\mu}$ interactions,
and 0.8\% were events caused by $\nu_{\mu}$ NC interactions.
The other sources of background contribute for less than 0.1\%.
\begin{table}[htbp]
	\caption{Number of selected events remaining after each selection check described in Sec.~\ref{sec:EVENT_SELECTION}. The purity indicates the fraction of  $\nu_{\mu}$ CC interactions in the MC prediction~(third column). Comparisons of the data with the MC prediction are only possible for numbers of neutrino-iron interactions which occurred inside the ECC bricks.}
	\label{tab:num_sel_events}
	\begin{center}
	\begin{tabular}{lccc}
		\hline
		\hline
		 Step & Data & MC & Purity \\
		\hline
		ECC--Shifter--INGRID track matching      & 9\,397  & - & - \\
		Fiducial volume cut                                         &    236   & - & - \\
		Manual microscope check                              &    203   & - & - \\
		Partner track search                                        &    202 & 207.6  & 81.7\% \\
		Kink event cut                                                  &    195 & 198.1  & 85.5\% \\
		Momentum consistency check                      &    183 & 188.8  & 88.2\% \\
		\hline
		\hline
	\end{tabular}
	\end{center}
\end{table}

\begin{figure}[htbp]
	\begin{center}
		\includegraphics[width=12cm]{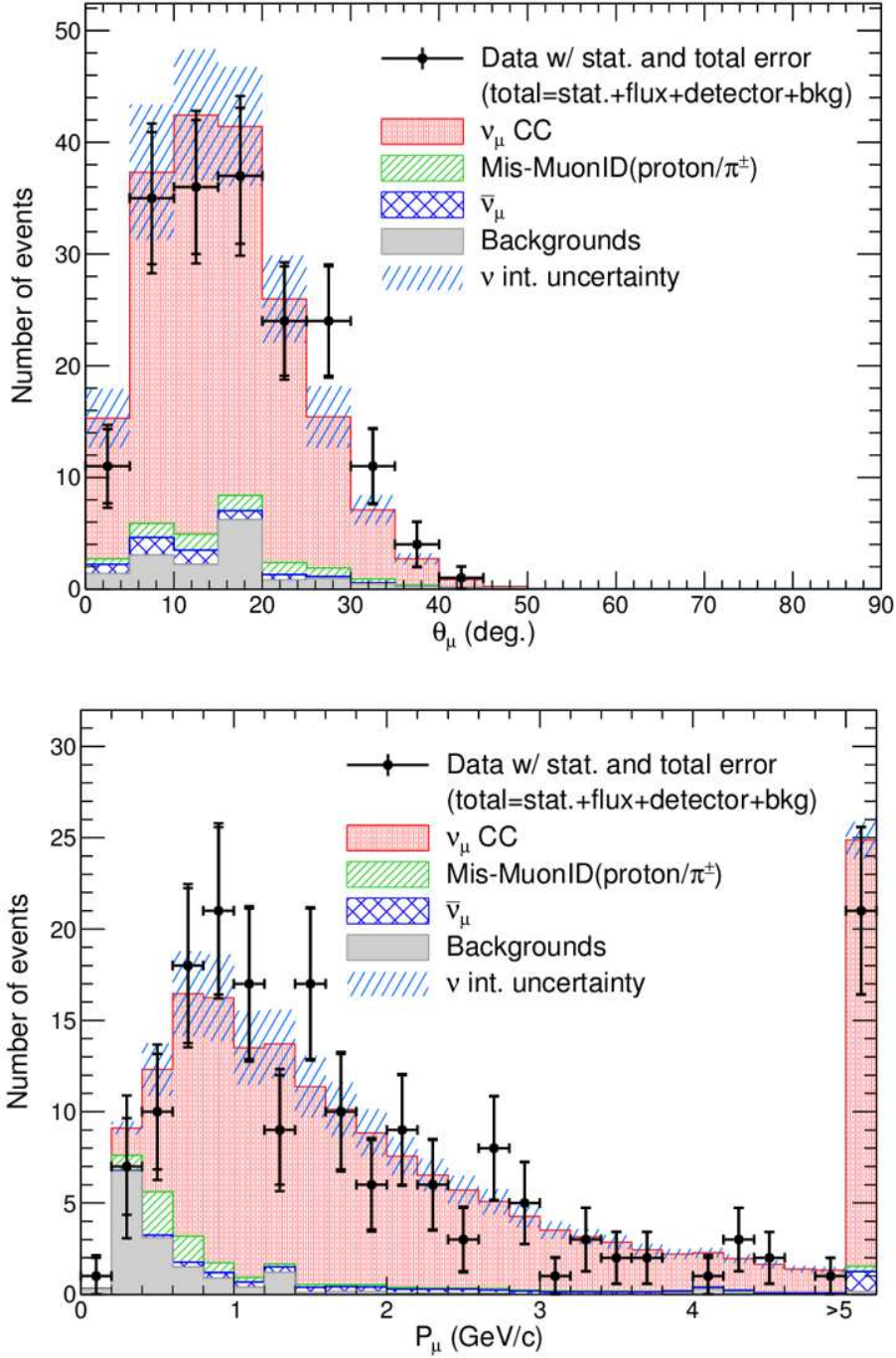}
		\caption{Distributions of muon kinematics from neutrino-iron interactions and backgrounds. The top figure shows the emission angle distribution, while the bottom figure shows the momentum distribution. In the right most bin of the momentum distribution, all the events with momenta above 5\,GeV/$c$ are contained. The data are shown by marker points and the MC predictions are shown by histograms. Inside error bars of the data represent statistical errors and outside error bars represent total errors, which are the quadrature sum of the statistical error and the uncertainties of the neutrino flux, the detector response, and the background estimation. Hatched regions of the MC predictions represent the uncertainties of the neutrino interaction model. The systematic uncertainties are described in Sec.~\ref{sec:SYSTEMATIC UNCERTAINTIES}.}
		\label{fig:muon_angle_momentum}
	\end{center}
\end{figure}

Figure~\ref{fig:muon_angle_momentum} shows the measured kinematics of the induced muons.
Here the emission angles of the induced muons are taken with respect to the neutrino beam direction.
To study the neutrino interaction models in comparison of the data and the MC prediction, the flux, detector response and background estimation uncertainties are included in the uncertainties on the data as well as the statistical uncertainty, while the uncertainties of the neutrino interaction models are included in the MC prediction uncertainty.
The systematic uncertainties are described in Sec.~\ref{sec:SYSTEMATIC UNCERTAINTIES}.
We found good agreements between the observed data and the MC predictions for the muon kinematic distributions.
The distributions show the reliability of our detector and data analysis.

\begin{figure}[htbp]
	\begin{center}
		\includegraphics[width=12cm]{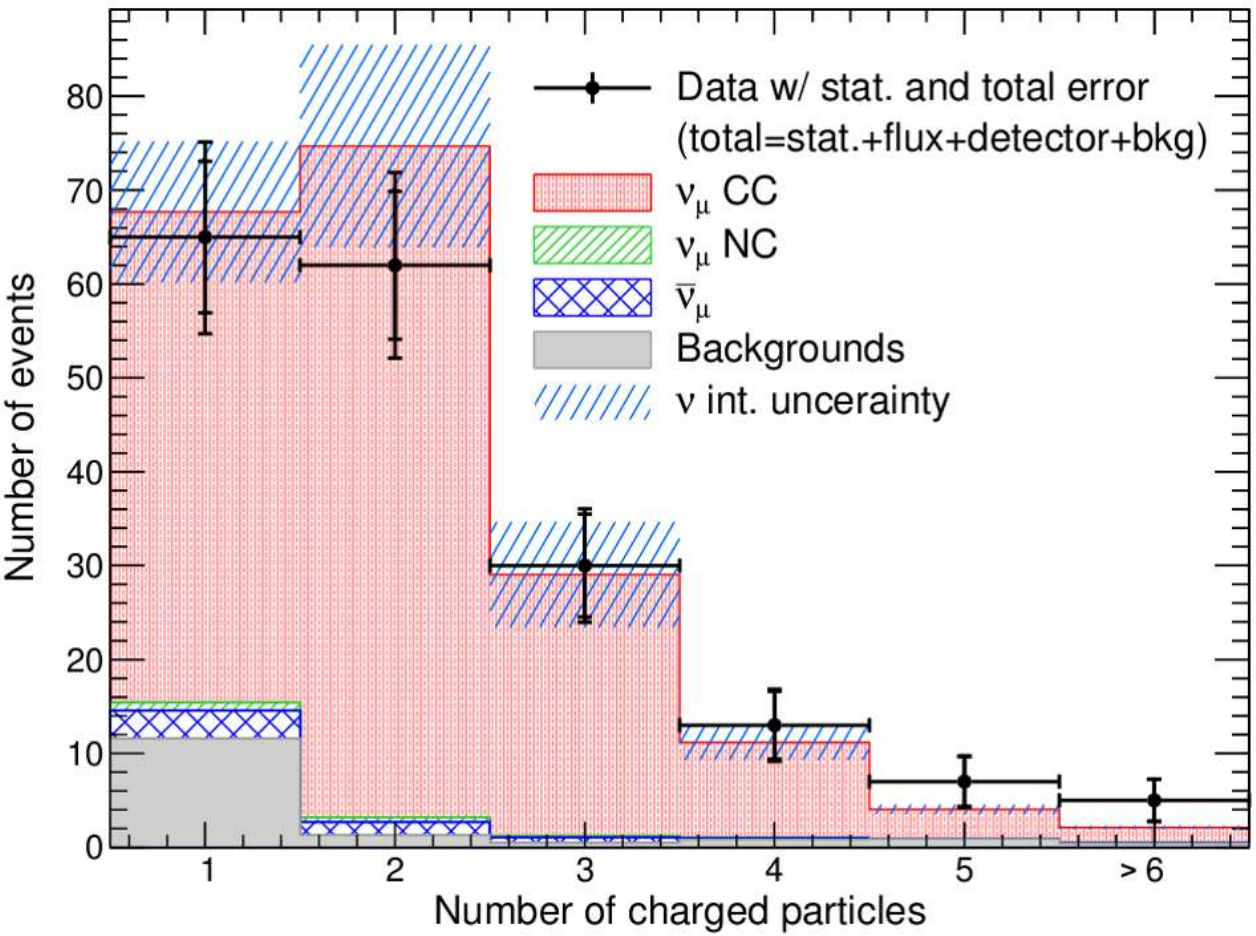}
		\caption{Multiplicity of the charged particles from neutrino-iron interactions including muon candidates. In the right most bin of the distribution, all the events above six-prong events are contained. The data are shown by marker points and the MC predictions are shown by histograms. Inside error bars of the data represent statistical errors and outside error bars represent total errors, which are the quadrature sums of the statistical error and the uncertainties of the neutrino flux, the detector response, and the background estimation. Hatched regions of the MC predictions represent the uncertainties of the neutrino interaction model. The systematic uncertainties are described in Sec.~\ref{sec:SYSTEMATIC UNCERTAINTIES}.}
		\label{fig:multiplicity}
	\end{center}
\end{figure}

Figure~\ref{fig:multiplicity} shows the multiplicity of charged particles from the neutrino-iron interactions.
The partner tracks, mainly protons and charged pions, are successfully detected by the emulsion detector in addition to muons although this analysis focuses only on muons for the CC-inclusive cross-section measurement.
These particles are expected to be detected with low momentum thresholds which are around 200\,MeV/$c$ for protons and 50\,MeV/$c$ for charged pions.
Although the statistical uncertainty is large, this measurement demonstrates the capability of the detailed study of the charged particles using the emulsion detector.
The multiplicity measurement is a basic study for exclusive channels, such as CC0$\pi$0p and CC0$\pi$1p, that will provide a better understanding of the neutrino interactions.

\begin{figure}[htbp]
	\begin{center}
		\includegraphics[width=10cm]{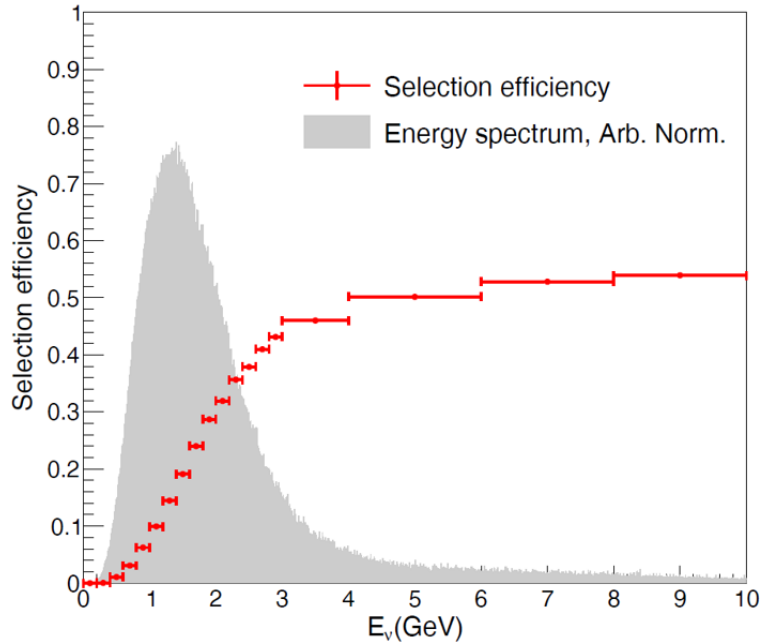}
		\caption{Selection efficiency of the $\nu_{\mu}$ CC interactions as a function of the neutrino energy. The selection efficiency was evaluated using the number of $\nu_{\mu}$ CC interactions in the FV as the denominator and the number of selected events as the numerator. The neutrino energy spectrum of the CC interactions in the FV is shown in gray.}
		\label{fig:detection_efficiency_numu_enu}
	\end{center}
\end{figure}

Figure~\ref{fig:detection_efficiency_numu_enu} shows the selection efficiency of the $\nu_{\mu}$ CC interactions as a function of the neutrino energy as estimated by the MC simulation.
The selection efficiency was defined using the number of $\nu_{\mu}$ CC interactions in the FV as the denominator and the number of selected events as the numerator.
The profile of selection efficiency curve is determined by the following factors: the connection efficiency between the ECC bricks and the Shifter, and between the ECC bricks and INGRID; the muon momentum threshold of the ECC--INGRID track matching.
The mean selection efficiency is 25.3\%.

\section{Cross-section measurement}\label{sec:ANALYSIS}
The flux-averaged $\nu_{\mu}$ CC inclusive cross section is measured from the number of selected events after background subtraction and efficiency correction.
The flux-averaged cross section is expressed as
\begin{equation}
\sigma_{\rm CC} = \frac{N_{\rm sel} - N_{\rm bkg}}{\phi T \varepsilon},
\label{eq:CC_total_cross_section}
\end{equation}
where $N_{\rm sel}$ is the number of events selected from the data, $N_{\rm bkg}$ is the number of background events predicted by the MC simulation, $T$ is the number of target nucleons in the FV, $\phi$ is the integrated $\nu_{\mu}$ flux for the ECC bricks, and $\varepsilon$ is the selection efficiency predicted by the MC simulation.
The cross section in a limited kinematic phase space of induced muons, $\theta_{\mu} < 45^{\circ}$ and $p_{\mu} > 400 \, {\rm MeV}/c$, is also expressed as Eq.~(\ref{eq:CC_total_cross_section}).
The values used in the cross-section measurements are summarized in Table~\ref{tab:CC_total_cross_section_measurement}.
\begin{table}[htbp]
	\caption{Cross-section measurement inputs.}
	\label{tab:CC_total_cross_section_measurement}
	\begin{center}
	\begin{tabular}{cccccc}
		\hline
		\hline
		Cross section & $N_{\rm sel}$ & $N_{\rm bkg}$ & $\phi$\,(cm$^{-2}$) & $T$\,(nucleons) & $\varepsilon$\,(\%) \\
		\hline
		$\sigma^{\mathrm{Fe}}_{\mathrm{CC}}$ &  183 & 22.3 & 1.94$\times$10$^{12}$ & 2.56$\times$10$^{28}$ & 25.3 \\
		$\sigma^{\mathrm{Fe}}_{\mathrm{CC \hspace{1mm} phase \hspace{0.5mm} space}}$ &  175 & 19.7 & 1.94$\times$10$^{12}$ & 2.56$\times$10$^{28}$ & 37.2 \\
		\hline
		\hline
	\end{tabular}
	\end{center}
\end{table}

\section{Systematic uncertainties}\label{sec:SYSTEMATIC UNCERTAINTIES}
The sources of systematic uncertainties in the cross-section measurement can be categorized into four groups: neutrino flux, neutrino interaction models, background estimation and detector response.
Each systematic uncertainty is described in Sec.~\ref{subsec:syst_flux}--\ref{subsec:syst_detector}.
The summary of the systematic uncertainties is described in Sec.~\ref{subsec:syst_summary}.

\subsection{Neutrino flux}\label{subsec:syst_flux}
\begin{figure}[htbp]
	\begin{center}
		\includegraphics[width=12cm]{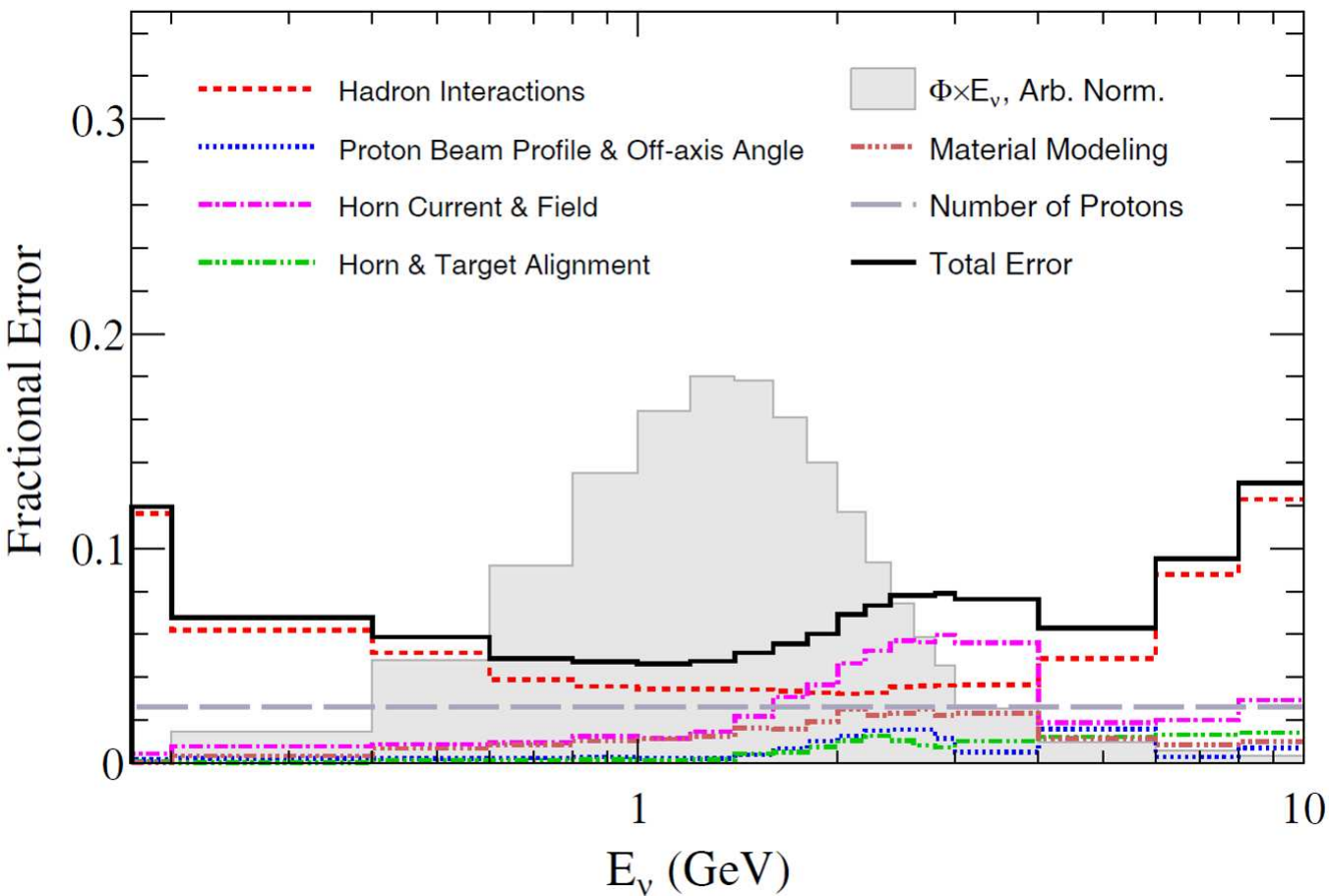}
		\caption{Neutrino flux uncertainties at the location of the detector arising from the hadron production uncertainties and the T2K beamline uncertainties.}
		\label{fig:flux_total_error_numubeam}
	\end{center}
\end{figure}

Uncertainties in the neutrino flux are due to uncertainties in hadron production and neutrino beam line optics.
A covariance matrix at the detector position was prepared following the same procedure as for the T2K experiment~\cite{Flux_JNUBEAM_2013}, with the relative errors in each energy bin shown in Fig.~\ref{fig:flux_total_error_numubeam}.
In the cross-section measurement, the neutrino flux is made to fluctuate according to the covariance matrix and the $\pm$1$\sigma$ change of the cross-section result is taken as the systematic uncertainty.
The cross-section uncertainty resulting from the flux uncertainties in the full~(restricted) phase space is found to be $-$5.8\%/$+$6.6\%~($-$5.9\%/$+$6.5\%).

\subsection{Neutrino interaction}\label{subsec:syst_interaction}
Table~\ref{tab:nu_int_prm_value_uncertainty} summarizes the parameters~\cite{T2K_Oscillation_2017,WAGSCI _2019} used for modeling the neutrino interactions and the FSI in NEUT.
It lists the nominal values and the $1 \sigma$ uncertainties for all the parameters.
These uncertainties affect the number of background events and the selection efficiency predicted by the MC simulation.
The combined uncertainty in the full~(restricted) phase space is $-$4.1\%/$+$4.6\%~($-$1.9\%/$+$2.0\%).
\begin{table}[htbp]
	\caption{Nominal parameter values and their uncertainties in the neutrino interaction models. Detailed descriptions of the parameters are given in Refs.~\cite{T2K_Oscillation_2017,WAGSCI _2019}.}
	\label{tab:nu_int_prm_value_uncertainty}
	\begin{center}
	\begin{tabular}{lll}
		\hline
		\hline
		 Parameter & Nominal value & Uncertainty(1$\sigma$) \\
		\hline
		${\textit{M}}^{\textrm{QE}}_{\textrm{A}}$   & 1.05\,GeV/$c^2$ & 0.20\,GeV/$c^2$\\
		${\textit{M}}^{\textrm{RES}}_{\textrm{A}}$  & 0.95\,GeV/$c^2$& 0.15\,GeV/$c^2$\\
		${\textit{C}}^{\textrm{A}}_{\textrm{5}}$(0)  & 1.01 & 0.12 \\
		Isospin $\frac{1}{2}$BG                           & 1.30 & 0.20 \\
		CC other shape                                            & 0 & 0.40 \\
		CC coherent normalization                       & 100$\%$ & 100$\%$ \\
		NC other normalization                             & 100$\%$ &   30$\%$ \\
		NC coherent normalization                       & 100$\%$ &   30$\%$ \\
		2p2h normalization                                     & 100$\%$ & 100$\%$ \\
		Fermi momentum ${\textit{P}}_{\textrm{F}}$                                      &  250\,MeV/$c$ & 30\,MeV/$c$ \\
		Binding energy ${\textit{E}}_{\textrm{b}}$                                             &     33\,MeV        &    9\,MeV \\
		Pion  absorption normalization                                                                     & 1.1 & 50$\%$ \\
		Pion  charge exchange normalization~($p_{\pi} <$ 500\,MeV/$c$)    & 1.0 & 50$\%$ \\
		Pion  charge exchange normalization~($p_{\pi} >$ 500\,MeV/$c$)    & 1.8 & 30$\%$ \\
		Pion  quasi elastic normalization~($p_{\pi} <$ 500\,MeV/$c$)           &  1.0 & 50$\%$ \\
		Pion  quasi elastic normalization~($p_{\pi} >$ 500\,MeV/$c$)           &  1.8 & 30$\%$ \\
		Pion  inelastic normalization                                                                          &  1.0 & 50$\%$ \\
		\hline
		\hline
	\end{tabular}
	\end{center}
\end{table}

\subsection{Background estimation}\label{subsec:syst_bkg_estimation}
For the background estimation, the uncertainties associated with the wall backgrounds and the misconnected backgrounds are considered.
The number of sand muons in the MC simulation was found to be 30\% smaller than in the data. 
To estimate the background, we used the observed number of sand muons and considered the 30\% difference as a systematic uncertainty which reflects our lack of understanding of the flux and the materials surrounding the detector hall.
The uncertainty attributed to misconnected events was evaluated using mock data, which are the combination of the nominal and fake data in which the time information of the ECC tracks is shifted.
This uncertainty is asymmetric because the misconnection rate of the beam-induced tracks and that of the cosmic-ray tracks are different.
The positive and negative uncertainties corresponding to the number of misconnected events were $+$24\% and $-$39\%, respectively.
The combined uncertainty in the full~(restricted) phase space is $-$1.8\%/$+$2.4\%~($-$1.1\%/$+$1.7\%), which is smaller than the other uncertainties.

\subsection{Detector response}\label{subsec:syst_detector}
The uncertainties associated with the detector response were estimated using the data and the MC simulation.
We considered the contributions of the following uncertainties: 
the detection efficiency of the base tracks, 
the track reconstruction in the ECC brick, 
the track connection between the ECC bricks, 
the track connection between the ECC bricks and the Shifter, 
the track matching between the ECC bricks and INGRID, 
the track reconstruction in the INGRID module, 
the kink cut, 
the momentum consistency check, 
the target mass, 
and the difference between iron and the stainless steel.
To evaluate the effect of the detector response uncertainties on the cross section, the MC simulations were run using each of the detector responses with their $1 \sigma$ uncertainty applied.
The difference between the cross-section result and its nominal result was defined as the systematic uncertainty for each detector response.
The effect on the cross-section measurement due to using the stainless steel plates instead of iron plates was estimated to be 0.3\%.
The main uncertainty components are the track matching between the ECC bricks and the Shifter, and the ECC bricks and INGRID, both of which were found to be 2--3\%.
In total, the uncertainty associated with the detector response on the cross-section result in the full~(restricted) phase space is $-$4.2\%/$+$4.4\%~($-$4.1\%/$+$4.2\%).

\subsection{Summary of the systematic uncertainties}\label{subsec:syst_summary}
Table~\ref{tab:xsec_systematic_uncertainty} summarizes the uncertainties involved in the cross-section measurements.
The total systematic uncertainty of the cross section measurement is estimated as a quadratic sum of the uncertainties of the neutrino flux, neutrino interaction, 	background estimation, and detector response.
The total systematic uncertainty involved in the cross-section measurement in the full~(restricted) phase space was $-$8.5\%/$+$9.4\%~($-$7.5\%/$+$8.2\%).
\begin{table}[htbp]
	\caption{Summary of the systematic uncertainties involved in the cross-section measurements.}
	\label{tab:xsec_systematic_uncertainty}
	\begin{center}
	\begin{tabular}{lcc}
		\hline
		\hline
		Item & $\sigma_{\mathrm{CC}}^{\mathrm{Fe}}$ & $\sigma^{\mathrm{Fe}}_{\mathrm{CC \hspace{1mm} phase \hspace{0.5mm} space}}$\\
		\hline
		Neutrino flux	                                                                                                 & $-$5.8$\%$ +6.6$\%$ & $-$5.9$\%$ +6.5$\%$ \\
		\hline
		${\textit{M}}^{\textrm{QE}}_{\textrm{A}}$                                      & $-$0.0$\%$ $+$1.5$\%$ & $-$0.0$\%$ +0.9$\%$ \\
		${\textit{M}}^{\textrm{RES}}_{\textrm{A}}$                                     & $-$0.0$\%$ $+$0.1$\%$ & $-$0.3$\%$ +0.2$\%$ \\
		${\textit{C}}^{\textrm{A}}_{\textrm{5}}$(0)                                     & $-$1.2$\%$ $+$1.1$\%$ & $-$0.7$\%$ +0.6$\%$ \\
		Isospin $\frac{1}{2}$BG                                                                            & $-$0.9$\%$ $+$0.8$\%$ & $-$0.3$\%$ +0.3$\%$ \\
		CC other shape                                                                                              & $-$0.6$\%$ $+$0.5$\%$ & $-$0.3$\%$ +0.2$\%$ \\
		CC coherent normalization                                                                         & $-$1.5$\%$ $+$1.6$\%$ & $-$0.7$\%$ +0.7$\%$ \\
		NC other normalization                                                                               & $-$1.0$\%$ $+$1.0$\%$ & $-$0.4$\%$ +0.4$\%$ \\
		NC coherent normalization                                                                         & $-$0.8$\%$ $+$0.0$\%$ & $-$0.2$\%$ +0.0$\%$ \\
		2p2h normalization                                                                                       & $-$2.5$\%$ $+$2.8$\%$ & $-$1.1$\%$ +1.2$\%$ \\
		Fermi momentum ${\textit{P}}_{\textrm{F}}$                                     & $-$1.1$\%$ $+$1.0$\%$ & $-$0.5$\%$ +0.4$\%$ \\
		Binding energy ${\textit{E}}_{\textrm{b}}$                                           & $-$0.9$\%$ $+$0.0$\%$ & $-$0.3$\%$ +0.2$\%$ \\
		Pion  absorption normalization                                                                   & $-$0.9$\%$ $+$1.0$\%$ & $-$0.4$\%$ +0.5$\%$ \\
		Pion  charge exchange normalization~($p_{\pi} <$ 500\,MeV/$c$)  & $-$0.0$\%$ $+$0.8$\%$ & $-$0.0$\%$ +0.2$\%$ \\
		Pion  charge exchange normalization~($p_{\pi} >$ 500\,MeV/$c$)  & $-$0.0$\%$ $+$0.8$\%$ & $-$0.0$\%$ +0.2$\%$ \\
		Pion  quasi elastic normalization~($p_{\pi} <$ 500\,MeV/$c$)         &  $-$0.8$\%$ $+$0.7$\%$ & $-$0.3$\%$ +0.2$\%$ \\
		Pion  quasi elastic normalization~($p_{\pi} >$ 500\,MeV/$c$)         &  $-$0.0$\%$ $+$0.8$\%$ & $-$0.2$\%$ +0.2$\%$ \\
		Pion  inelastic normalization                                                                        &  $-$0.8$\%$ $+$0.7$\%$ & $-$0.3$\%$ +0.2$\%$ \\
		\hline
		Wall backgrounds                                                                                            &  $-$1.1$\%$ $+$1.1$\%$ & $-$0.2$\%$ +0.2$\%$ \\
		ECC--Shifter--INGRID misconnection backgrounds                               &  $-$1.4$\%$ $+$2.2$\%$ & $-$1.1$\%$ +1.7$\%$ \\
		\hline
		Base track detection efficiency                                                                      &  $-$0.3$\%$ $+$0.1$\%$ & $-$0.3$\%$ +0.1$\%$ \\
		ECC track reconstruction                                                                               &  $-$0.1$\%$ $+$0.1$\%$ & $-$0.1$\%$ +0.1$\%$ \\
		ECC bricks track connection                                                                          &  $-$0.1$\%$ $+$0.1$\%$ & $-$0.1$\%$ +0.1$\%$ \\
		ECC--Shifter track connection                                                                      &  $-$2.3$\%$ $+$2.4$\%$ & $-$2.3$\%$ +2.3$\%$ \\
		ECC--INGRID track connection                                                                   &  $-$3.0$\%$ $+$3.2$\%$ & $-$3.1$\%$ +3.2$\%$ \\
		INGRID track reconstruction                                                                        &  $-$0.7$\%$ $+$0.8$\%$ & $-$0.7$\%$ +0.8$\%$ \\
		Kink event cut                                                                                                   &  $-$0.6$\%$ $+$0.5$\%$ & $-$0.2$\%$ +0.1$\%$ \\
		Momentum consistency check                                                                        &  $-$1.3$\%$ $+$1.3$\%$ & $-$0.8$\%$ +0.8$\%$ \\
		Target mass                                                                                                       &  $-$0.6$\%$ $+$0.6$\%$ & $-$0.7$\%$ +0.7$\%$ \\
		Difference between iron and the stainless steel                                          &  $-$0.3$\%$ $+$0.3$\%$ & $-$0.3$\%$ +0.3$\%$ \\
		\hline
		Total                                                                                                                    &  $-$8.5$\%$ $+$9.4$\%$ & $-$7.5$\%$ +8.2$\%$ \\
		\hline
		\hline
	\end{tabular}
	\end{center}
\end{table}

\section{Results and discussion}\label{sec:RESULTS}
\begin{figure}[htbp]
	\begin{center}
		\includegraphics[width=12cm]{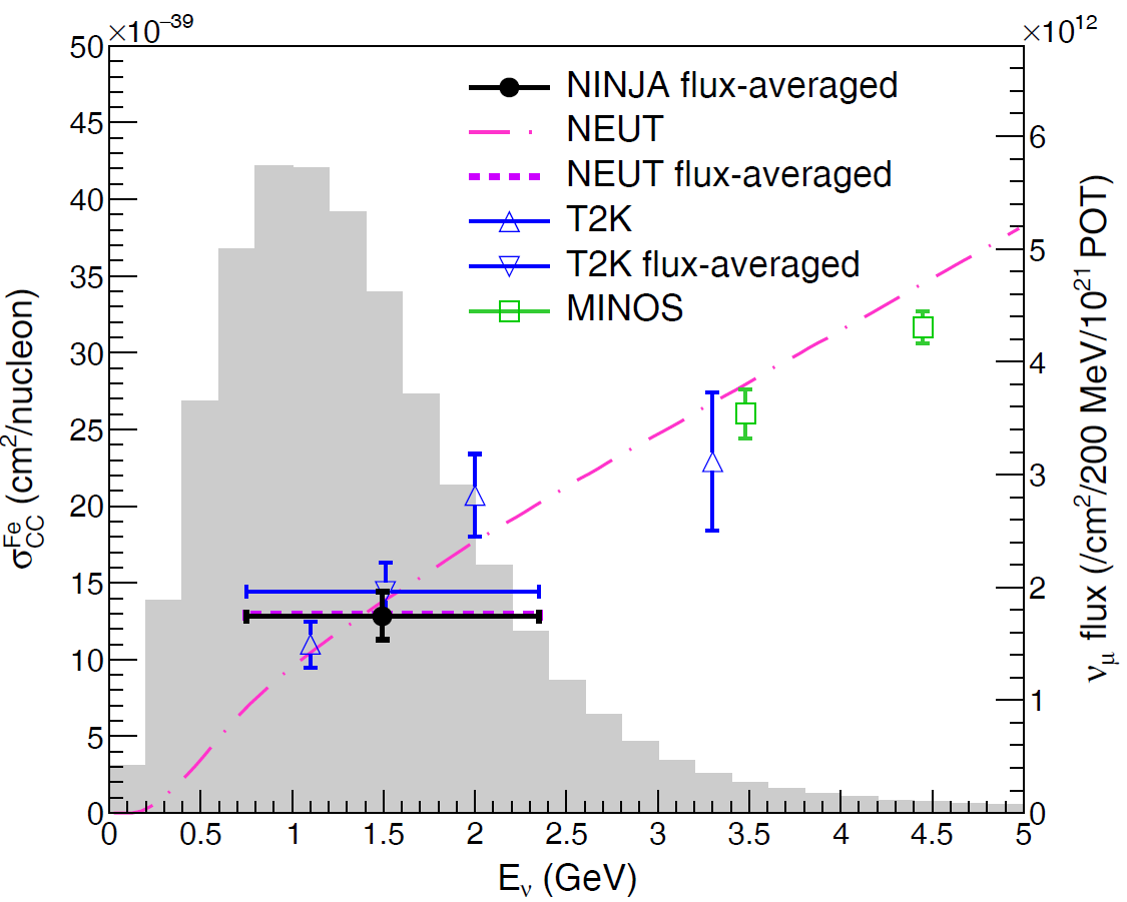}
		\caption{Flux-averaged $\nu_{\mu}$ CC inclusive cross section on iron. Our data point is plotted at the mean flux energy. The vertical error bar represents the total~(statistical and systematic) uncertainty and the horizontal bar represents 68\% of the flux at each side of the mean energy. The MINOS and T2K results are also plotted. The MINOS results~\cite{MINOS_CCinclusive_Xsec_2010} are the $\nu_{\mu}$ CC inclusive cross sections on iron. The T2K results are the flux-averaged $\nu_{\mu}$ CC inclusive cross section~\cite{INGRID_CCinclusive_Xsec_2014} and the $\nu_{\mu}$ CC inclusive cross sections~\cite{INGRID_CCinclusive_Xsec_2016} on iron. The neutrino flux at the detector position is shown in gray.}
		\label{fig:Xsec_meas_numu}
	\end{center}
\end{figure}

The measured flux-averaged $\nu_{\mu}$ CC inclusive cross section on iron is 
\begin{equation}
\sigma^{\mathrm{Fe}}_{\mathrm{CC}} = (1.28 \pm 0.11({\mathrm{stat.}})^{+0.12}_{-0.11}({\mathrm{syst.}})) \times 10^{-38} \,  {\mathrm{cm}}^{2}/{\mathrm{nucleon}}, 
\label{eq:CC_total_cross_section_result}
\end{equation}
at a mean neutrino energy of 1.49\,GeV.
This is the cross section per nucleon for an iron nucleus.
Figure~\ref{fig:Xsec_meas_numu} shows the cross-section result obtained in this study as well as those reported by other experiments.
The result measured by the emulsion-based detector is consistent with the T2K measurements using INGRID on the same beamline $(\, (1.444 \pm 0.002({\rm stat.}) ^{+0.189}_{-0.157} ({\rm syst.})) \times 10^{-38} \, {\rm cm^{2}}/{\rm nucleon} \,)$~\cite{INGRID_CCinclusive_Xsec_2014,INGRID_CCinclusive_Xsec_2016}, although a part of the systematic uncertainty is expected to be correlated between the two measurements.
The measured cross section agrees well with the MC prediction of 1.30$\times$10$^{-38}$\,cm$^{2}$.
The cross section for a restricted phase space of induced muons, $\theta_{\mu} < 45^{\circ}$ and $p_{\mu} > 400 \, {\rm MeV}/c$, on iron is
\begin{equation}
\sigma^{\mathrm{Fe}}_{\mathrm{CC \hspace{1mm} phase \hspace{0.5mm} space}} = (0.84 \pm 0.07({\mathrm{stat.}})^{+0.07}_{-0.06}({\mathrm{syst.}})) \times 10^{-38} \,  {\mathrm{cm}}^{2}/{\mathrm{nucleon}}. 
\label{eq:CC_total_cross_section_phase_space_result}
\end{equation}
This result is also consistent with the T2K measurement using INGRID for the same phase space $(\, (0.859 \pm 0.003({\rm stat.}) ^{+0.12}_{-0.10} ({\rm syst.})) \times 10^{-38} \, {\rm cm^{2}}/{\rm nucleon} \,)$~\cite{WAGSCI _2019}, although some correlations in the systematic uncertainty are expected.
It is also in good agreement with the MC prediction of 0.87$\times$10$^{-38}$\,cm$^{2}$.
This cross-section measurement was performed via different techniques from the T2K measurement.
The systematic uncertainty on this measurement was reduced owing to the improved flux prediction based on recent hadron-production data~\cite{NA61_t2k_replica_target_2013,NA61_t2k_replica_target_2016,NA61_t2k_thin_target_2011,NA61_t2k_thin_target_2012,NA61_t2k_thin_target_2014,NA61_t2k_thin_target_2016}, although the statistical uncertainty is large.
These results also show the reliability of our detector and validate our understanding on the whole chain of the data analysis.

\section{Conclusions}\label{sec:CONCLUSIONS}
In this study, we report the first $\nu_{\mu}$ cross-section measurement using an iron-target emulsion detector in the NINJA pilot experiment.
The measurement was performed by exposing a 65\nobreakdash-kg iron target to the neutrino beam at J\nobreakdash-PARC.
From the data acquired during the 4.0$\times$10$^{19}$ POT exposure, the flux-averaged $\nu_{\mu}$ CC inclusive cross sections on iron at a mean neutrino energy of 1.49\,GeV were measured.
We have reported the measurement of cross sections in the full phase space and a limited phase space for the kinematics of the induced muons with $\theta_{\mu} < 45^{\circ}$ and $p_{\mu} > 400 \, {\rm MeV}/c$.
The results of the cross-section measurement are consistent with the T2K measurements as well as the current neutrino interaction models.
We also found good agreements between the observed data and the MC predictions for the muon kinematic distributions. 
These results demonstrate that we have a good understanding of neutrino interactions around 1\,GeV.
In addition, the feasibility of the emulsion detector for the neutrino cross-section measurement was demonstrated.
The emulsion detector provides sub-$\mu$m spatial resolution and a low momentum threshold for the protons and charged pions produced in neutrino interactions.
Precise measurements of the kinematics of these particles and cross-section measurements for exclusive channels, such as CC0$\pi$0p and CC0$\pi$1p, will provide a better understanding of the neutrino interactions in the near future.
The charged particles from the neutrino-iron interactions are successfully detected, and their multiplicities are measured using the emulsion detector.
This measurement demonstrates the capability of neutrino-nucleus interaction measurements with the emulsion detector.
Our future work will conduct a detailed study of the neutrino interactions in the 1\,GeV energy region using emulsion detectors.
This study will be important for future long-baseline neutrino oscillation experiments.

\section*{Acknowledgments}
We would like to acknowledge the support of the T2K Collaboration in performing the experiment.
Furthermore, we appreciate the assistance of the T2K neutrino beam group in providing a high-quality beam and the MC simulation.
We thank the T2K INGRID group for providing access to their data.
We acknowledge the work of the J\nobreakdash-PARC staff in facilitating superb accelerator performance.
We would like to thank P. Vilain~(Brussels University, Belgium) for his careful reading of the manuscript and for his valuable comments.
This work was financially supported by the Japan Society for the Promotion of Science~(JSPS) KAKENHI Grant Numbers JP25105001, JP25105006, JP26105516, JP26287049, JP25707019, JP20244031, JP26800138, JP16H00873, JP18K03680, JP17H02888, JP18H03701, JP18H05537, and JP18H05541.
%



\begin{thebibliography}{99}

\bibitem{T2K_2011}
K. Abe $\textit{et al.}$ (T2K Collaboration), Nucl. Instrum. Meth. Phys. Res. A \textbf{659}, 106--135 (2011).
\bibitem{NINJA_Run4_ECC_2017}
T. Fukuda $\textit{et al.}$, Prog. Theor. Exp. Phys. \textbf{2017}, 063C02 (2017).
\bibitem{NINJA_Run4_Shifter_2017}
K. Yamada $\textit{et al.}$, Prog. Theor. Exp. Phys. \textbf{2017}, 063H02 (2017).
\bibitem{NINJA_Run8_2020}
A. Hiramoto $\textit{et al.}$ (NINJA Collaboration), Phys. Rev. D \textbf{102}, 072006 (2020).

\bibitem{MicroBooNE_multiplicity_2019}
C. Adams $\textit{et al.}$ (MicroBooNE Collaboration), Eur. Phys. J. C \textbf{79}, 248 (2019).
\bibitem{T2K_proton_2018}
K. Abe $\textit{et al.}$ (T2K Collaboration), Phys. Rev. D \textbf{98}, 032003 (2018).
\bibitem{MINERvA_CCQE_2013}
G. A. Fiorentini $\textit{et al.}$ (MINERvA Collaboration), Phys. Rev. Lett. \textbf{111}, 022502 (2013).





\bibitem{S.Takahashi_2010}
S. Takahashi $\textit{et al.}$, Nucl. Phys. A \textbf{620}, 192--195 (2010).
\bibitem{S.Takahashi_2016}
S. Takahashi $\textit{et al.}$ (GRAINE Collaboration), Prog. Theor. Exp. Phys. \textbf{2016}, 073F01 (2016).
\bibitem{mizutani_2019}
F. Mizutani, Ph.D. thesis, Kobe University (2019) (in Japanese).
\bibitem{INGRID_2010}
K. Abe $\textit{et al.}$ (T2K Collaboration), Nucl. Instrum. Meth. Phys. Res. A \textbf{694}, 211--223 (2012).

\bibitem{Flux_JNUBEAM_2013}
K. Abe $\textit{et al.}$ (T2K Collaboration), Phys. Rev. D \textbf{87}, 012001 (2013).




\bibitem{morimoto_2020}
Y. Morimoto, Ph.D. thesis, Toho University (2020) (in Japanese).
\bibitem{nishio_2020}
A. Nishio $\textit{et al.}$, Nucl. Instrum. Meth. Phys. Res. A \textbf{966}, 163850 (2020).

\bibitem{FTS_2013} 
T. Fukuda $\textit{et al.}$, J. Instrum. \textbf{8}, P01023 (2013).
\bibitem{FTS_2014} 
T. Fukuda $\textit{et al.}$, J. Instrum. \textbf{9}, P12017 (2014).


\bibitem{K.Kodama_2006}
K. Kodama $\textit{et al.}$, Adv. Space Res. \textbf{37}, 2120--2124 (2006).
\bibitem{S.Takahashi_2018}
S. Takahashi and S. Aoki $\textit{et al.}$ (GRAINE Collaboration), Adv. Space Res. \textbf{62}, 2945--2953 (2018).




\bibitem{NEUT_2002}
Y. Hayato, Nucl. Phys. B Proc. Suppl. \textbf{112}, 171--176 (2002).
\bibitem{NEUT_2009} 
Y. Hayato, Acta Phys. Pol. B \textbf{40}, 2477--2489 (2009).
\bibitem{Geant4_2003}
S. Agostineli $\textit{et al.}$, Nucl. Instrum. Meth. Phys. Res. A \textbf{506}, 250--303 (2003).
\bibitem{Geant4_2006}
J. Allison $\textit{et al.}$, IEEE Trans. Nucl. Sci. \textbf{53}, 270--278 (2006).
\bibitem{Geant4_2016}
J. Allison $\textit{et al.}$, Nucl. Instrum. Meth. Phys. Res. A \textbf{835}, 186--225 (2016).


\bibitem{Geant_1994}
R. Brun, F. Carminati, and S. Giani, Report No. CERNW5013 (1994).
\bibitem{FLUKA_2005}
A. Ferrari, P. R. Sala, A. Fasso, and J. Ranft, Report No. CERN-2005-010; Report No. SLAC-R-773; Report No. INFN-TC-05-11 (2005).

\bibitem{FLUKA_2014}
T. Bohlen, F. Cerutti, M. Chin, A. Fasso, A. Ferrari, P. Ortega, A. Mairani, P. Sala, G. Smirnov, and V. Vlachoudis, Nucl. Data Sheets \textbf{120}, 211--214 (2014).
\bibitem{NA61_2014}
N. Abgrall $\textit{et al.}$ (NA61/SHINE Collaboration), J. Instrum. \textbf{9}, P06005 (2014).
\bibitem{NA61_t2k_replica_target_2013}
N. Abgrall $\textit{et al.}$ (NA61/SHINE Collaboration), Nucl. Instrum. Meth. Phys. Res. A \textbf{701}, 99--114 (2013).
\bibitem{NA61_t2k_replica_target_2016}
N. Abgrall $\textit{et al.}$ (NA61/SHINE Collaboration), Eur. Phys. J. C \textbf{76}, 617 (2016).
\bibitem{NA61_t2k_thin_target_2011}
N. Abgrall $\textit{et al.}$ (NA61/SHINE Collaboration), Phys. Rev. C \textbf{84}, 034604 (2011).
\bibitem{NA61_t2k_thin_target_2012}
N. Abgrall $\textit{et al.}$ (NA61/SHINE Collaboration), Phys. Rev. C \textbf{85}, 035210 (2012).
\bibitem{NA61_t2k_thin_target_2014}
N. Abgrall $\textit{et al.}$ (NA61/SHINE Collaboration), Phys. Rev. C \textbf{89}, 025205 (2014).
\bibitem{NA61_t2k_thin_target_2016}
N. Abgrall $\textit{et al.}$ (NA61/SHINE Collaboration), Eur. Phys. J. C \textbf{76}, 84 (2016).


\bibitem{SK_2003}
S. Fukuda $\textit{et al.}$ (Super-Kamiokande Collaboration), Nucl. Instrum. Meth. Phys. Res. A \textbf{501}, 418--462 (2003).
\bibitem{1p1h_nieves_2012}
J. Nieves, I. Ruiz Simo, and M. J. Vicente Vacas, Phys. Lett. B \textbf{707}, 72--75 (2012).
\bibitem{rpa_nieves_2012}
J. Nieves, J. E. Amaro, and M. Valverde, Phys. Rev. C \textbf{70}, 055503 (2004).
\bibitem{2p2h_nieves_2011}
J. Nieves, I. Ruiz Simo, and M. J. Vicente Vacas, Phys. Rev. C \textbf{83}, 045501 (2011).
\bibitem{res_rein_sehgal_1981}
D. Rein and L. M. Sehgal, Ann. Phys. (N. Y.) \textbf{133}, 79--153 (1981).
\bibitem{coh_rein_sehgal_1983}
D. Rein and L. M. Sehgal, Nucl. Phys. B \textbf{223}, 29--44 (1983).
\bibitem{coh_rein_sehgal_2007}
D. Rein and L. M. Sehgal, Phys. Lett. B \textbf{657}, 207--209 (2007).
\bibitem{dis_pdf_1998}
M. Gl$\ddot{\rm u}$ck, E. Reya, and A. Vogt, Eur. Phys. J. C \textbf{5}, 461--470 (1998).
\bibitem{dis_pdf_2003}
A. Bodek and U. K. Yang, AIP Conf. Proc. \textbf{670}, 110--117 (2003).
\bibitem{dis_pdf_2005}
A. Bodek, I. Park, and U. K.Yang, Nucl. Phys. Proc. Suppl. \textbf{139}, 113--118 (2005).
\bibitem{pion_fsi_2014}
P. de Perio, Ph.D. thesis, University of Toronto (2014).
\bibitem{pion_fsi_2019}
E. S. Pinzon Guerra $\textit{et al.}$, Phys. Rev. D \textbf{99}, 052007 (2019).


\bibitem{QGSPBERT_2009}
J. Apostolakis $\textit{et al.}$, J. Phys. Conf. Ser. \textbf{160}, 012073 (2009).

\bibitem{VPH_2006}
T. Toshito $\textit{et al.}$, Nucl. Instrum. Meth. Phys. Res. A \textbf{556}, 482--489 (2006).




\bibitem{HTS_2017}
M. Yoshimoto $\textit{et al.}$, Prog. Theor. Exp. Phys. \textbf{2017}, 103H01 (2017).
\bibitem{PH_2004}
T. Toshito $\textit{et al.}$, Nucl. Instrum. Meth. Phys. Res. A \textbf{516}, 436--439 (2004).
\bibitem{DOUNUT_2002}
K. Kodama $\textit{et al.}$ (DONUT Collaboration), Nucl. Instrum. Meth. Phys. Res. A \textbf{493}, 45--66 (2002).
\bibitem{NETSCAN2_2012}
K. Hamada $\textit{et al.}$ (OPERA Collaboration), J. Instrum. \textbf{7}, P07001 (2012).


\bibitem{INGRID_CCinclusive_Xsec_2014} 
K. Abe $\textit{et al.}$ (T2K Collaboration), Phys. Rev. D \textbf{90}, 052010 (2014).
\bibitem{INGRID_CCinclusive_Xsec_2016} 
K. Abe $\textit{et al.}$ (T2K Collaboration), Phys. Rev. D \textbf{93}, 072002 (2016).
\bibitem{cellular_automaton_2005}
H. Maesaka, Ph.D. thesis, Kyoto University (2005).




\bibitem{ScanBack_1979}
H. Fuchi $\textit{et al.}$, J. Phys. Soc. Jap. \textbf{47}, 687--694 (1979).
\bibitem{ScanBack_1997}
E. Eskut $\textit{et al.}$ (CHORUS Collaboration), Nucl. Instrum. Meth. Phys. Res. A \textbf{401}, 7--44 (1997).
\bibitem{ScanBack_2013}
J. Yoshida $\textit{et al.}$, J. Instrum. \textbf{8}, P02009 (2013).
\bibitem{DOUNUT_2001}
K. Kodama $\textit{et al.}$ (DONUT Collaboration), Phys. Lett. B \textbf{504}, 218--224 (2001).

\bibitem{NINJA_Run6_TAUP_2020}
H. Oshima $\textit{et al.}$, J. Phys. Conf. Ser. \textbf{1468}, 012128 (2020).

\bibitem{MCS_angular_method_2012}
N. Agafonova $\textit{et al.}$ (OPERA Collaboration), New J. Phys. \textbf{14}, 013026 (2012).

\bibitem{MCS_angular_coordinate_method_2007}
K. Kodama $\textit{et al.}$ (DONUT Collaboration), Nucl. Instrum. Meth. Phys. Res. A \textbf{574}, 192--198 (2007).



\bibitem{T2K_Oscillation_2017} 
K. Abe $\textit{et al.}$ (T2K Collaboration), Phys. Rev. D \textbf{96}, 092006 (2017).
\bibitem{WAGSCI _2019} 
K. Abe $\textit{et al.}$ (T2K Collaboration), Prog. Theor. Exp. Phys. \textbf{2019}, 093C02 (2019).

\bibitem{MINOS_CCinclusive_Xsec_2010} 
P. Adamson $\textit{et al.}$ (MINOS Collaboration), Phys. Rev. D \textbf{81}, 072002 (2010).


\end{thebibliography}
\end{document}